\documentclass{IEEEtran}
\usepackage{cite}
\usepackage{amsmath,amssymb,amsfonts}
\usepackage{algorithmic}
\usepackage{graphicx}
\usepackage{textcomp}
\def\BibTeX{{\rm B\kern-.05em{\sc i\kern-.025em b}\kern-.08em
		T\kern-.1667em\lower.7ex\hbox{E}\kern-.125emX}}
\usepackage{amsmath, bm}
\usepackage{amssymb}
\usepackage{booktabs}
\usepackage{multirow}
\usepackage{siunitx}
\usepackage{makecell}
\usepackage{balance}

\ifCLASSOPTIONcompsoc
\usepackage[caption=false,font=normalsize,labelfont=sf,textfont=sf]{subfig}
\else
\usepackage[caption=false,font=footnotesize]{subfig}
\fi

\begin{document}
	
\title{Fast Electromagnetic Validations of Large-Scale Digital Coding Metasurfaces Accelerated by \\Recurrence Rebuild and Retrieval Method}

\author{Yu Zhao, \IEEEmembership{Member, IEEE}, Shang Xiang, \IEEEmembership{Member, IEEE}, and Long Li, \IEEEmembership{Senior Member, IEEE}
	\thanks{This work has been submitted to the IEEE for possible publication. Copyright may be transferred without notice, after which this version may no longer be accessible. (Corresponding author: Yu Zhao and Long Li)}
	\thanks{Yu Zhao is with the School of Electronic Engineering, Xidian University, Xi’an 710071, China. (e-mail: zhaoyu775885@163.com).}
	\thanks{Long Li is with the Key Laboratory of High Speed Circuit Design and EMC, Ministry of Education, School of Electronic Engineering, Xidian University, Xi’an 710071, China. (e-mail: lilong@mail.xidian.edu.cn).}
	\thanks{Shang Xiang is with the Lund University, 22100 Lund, Sweden. (e-mail: bingoxiang917@163.com)}}

\maketitle

\begin{abstract}
The recurrence rebuild and retrieval method (R3M) is proposed in this paper to accelerate the electromagnetic (EM) validations of large-scale digital coding metasurfaces (DCMs).
R3M aims to accelerate the EM validations of DCMs with varied codebooks,
which involves the analysis of a group of similar but not identical structures.
The method transforms general DCMs to rigorously periodic arrays by replacing each coding unit with the macro unit,
which comprises all possible coding states.
The system matrix corresponding to the rigorously periodic array is globally shared for DCMs with arbitrary codebooks via implicit retrieval.
The discrepancy of the interactions for edge and corner units are precluded by the basis extension of periodic boundaries.
Moreover, the hierarchical pattern exploitation (HPE) algorithm is leveraged to efficiently assemble the system matrix for further acceleration.
Due to the fully utilization of the rigid periodicity,
the computational complexity of R3M-HPE is theoretically lower than that of $\mathcal{H}$-matrix within the same paradigm.
Numerical results for two types of DCMs indicate that R3M-HPE is accurate in comparison with commercial software.
Besides, R3M-HPE is also compatible with the preconditioning for efficient iterative solutions.
The efficiency of R3M-HPE for DCMs outperforms the conventional fast algorithms in both the storage and CPU time cost.

\end{abstract}


\section{Introduction}
\label{sec:introduction}
\IEEEPARstart{D}{igital} coding metasurface (DCM) has recently attracted considerable attention for the powerful capability of manipulating electromagnetic (EM) waves \cite{2014_Cui}.
As the 2-D counterpart of metamaterials, 
DCMs characterize the meta-atoms as quantized codes, 
providing a simple and efficient way to design the metasurface \cite{2017_AOM, 2021_NC_WANG, 2019_OE}.
Typical applications of DCM include the control of reflection and polarization of EM waves \cite{2012_APM_DRSMITH}, microwave imaging \cite{2020_TAP_Imaging}, 
radar cross-section reduction \cite{2020_TAP_RCS,PIERS_RCS_REDUCTION_ANT,PIERS_RCS_REDUCTION}, etc.
While the syntheses of the codebooks are efficient,
the numerical results are so rough that full-wave simulations are inevitable to validate the design, 
as is about to be shown in Section \ref{sec:validation}.
With the increasingly wide applications of large-scale DCMs,
the huge amount of design and validation tasks urgently require both accurate and fast solutions. 

Integral equations (IEs) based on the method of moments (MoM) are suitable solutions for analyzing the radiation and scattering of DCMs\cite{1998_TAP_MOM}.
Due to the prohibitive resource occupation of MoM, 
various fast algorithms have been developed to improve the computational efficiency and reduce the storage requirements in tackling large-scale problems.
Typical fast algorithm for IEs are mainly classified into three types \cite{2021_OJAP}.
The multiple expansion based methods, 
such as the fast multipole method (FMM) and the multilevel fast multipole algorithm (MLFMA), 
rely on the analytical harmonic expansion of the Green’s function \cite{1997_TAP_MLFMA}.
The fast Fourier transform (FFT) based methods \cite{Vladimir_2009_MTT, YangKai_2013_GRS, 2007_TAP_IEFFT} exploit the projection and interpolation of the basis functions onto regular grids 
and the convolutional form of the Green’s function 
to realize efficient solutions.
The low-rank factorization based methods, 
including the adaptive cross approximation (ACA) \cite{JinfaLee_2005_EMC}, 
fast nested cross approximation \cite{2019_TMTT_FNCA},
$\mathcal{H}$-matrix \cite{Hackbusch_1999} and $\mathcal{H}^2$-matrix \cite{2021_TAP_H2},
reveal the rank-deficient nature of coupling between two separated groups. 
Compared with the other two types of methods, the low-rank factorization based methods are usually kernel independent and thereby more flexible \cite{2021_OJAP}.

Though quite efficient for general problems, 
the aforementioned fast algorithms are still limited in performance when analyzing periodic arrays \cite{2010_TAP_CBF_AIM}.
Since the periodicity is special regularization,
numerical analyses of periodic arrays should be more efficient.
Several works have been proposed to deal with this problem.
The array decomposition method \cite{2003_TAP_ADM} takes the advantages of block-Toeplitz with hybrid finite element-boundary integral method,
leading to a linear storage complexity and $\mathcal{O}(N\log{}N)$ time complexity with the FFT-based iterative solution.
The two-level characteristic basis function method \cite{2012_TAP_CBF_FMM_FFT}, which is based on the plane wave derivation and local interaction, was further proposed with the acceleration of FMM-FFT .
The finite-element method with domain decomposition can utilize the geometrical repetitions of large
periodic structures to achieve fast parallel solution \cite{2020_AWPL_DDM}.
It is preferable to maintain the high efficiency of analyzing periodic arrays in the context of DCMs, which is of the quasi-periodic planar structure.
However, the above methods demand that the array elements be strictly identical and coded with the same state, 
which is a rare case for DCMs.



In this paper, the recurrence rebuild and retrieval method (R3M) is proposed to accelerate the validations of DCMs with varied codebooks. 
The method enables the construction of a global matrix, which corresponds to a rigorously periodic array that contains all possible codebooks.
The global matrix is assembled only once and shared to match arbitrary codebooks via the implicit retrieval.
The discrepancy of the interactions 
for edge and corner units are precluded with the basis 
extension of periodic boundaries. 
The assembly of the global matrix 
is further accelerated by the hierarchical pattern exploitation 
(HPE) algorithm, 
which was first proposed in \cite{2021_TAP_HPE} for the analysis of periodic arrays with high efficiency.
Owing to the rigid periodicity,
the computational complexity of R3M-HPE is theoretically lower than that of the classical $\mathcal{H}$-matrix within the same paradigm.
Numerical examples indicate that R3M-HPE is accurate in comparison with commercial software.
R3M-HPE is also compatible with the preconditioning for fast convergence.
The performance of R3M-HPE in the storage and CPU time cost outperform both $\mathcal{H}$-matrix and MLFMA , validating the high efficiency of the proposed method.

Our contribution is twofold.
First, the recurrence rebuild transforms general DCMs to rigorously periodic arrays,
which achieves the full potential of HPE for further acceleration.
Second, the implicit retrieval of the global matrix corresponding to the periodic array 
can flexibly adjust to arbitrary codebook.
The immediate merit is that the matrix assembly is executed merely once for validations of DCMs with varied codebooks.

The rest of this paper is organized as follows. 
In Section \ref{sec:validation}, we introduce the background of this work.
In Section \ref{sec:methodology}, 
we present the initial motivation 
and the basic concept of the recurrence rebuild and retrieval method.
The HPE algorithm is then briefly introduced as enhancement of R3M to fully exploit the periodicity.
The preconditioner and complexity analysis are also presented in Section \ref{sec:methodology}.
The performance of the proposed method, including the accuracy, convergency and efficiency, are verified with several numerical examples in Section \ref{sec:examples}.
The conclusions are drawn in Section \ref{sec:conclustion}.

\section{EM Validations of DCMs}
\label{sec:validation}
\subsection{Method of Moments Based EM Validations}
The DCMs are simultaneously depicted by the the state and topology of each coding unit.
The codebooks, 
which indicate the distribution of the coding states,
can be synthesized via the generalized Snell's law \cite{2011_snell} or the array theory \cite{1997_array}.
However, the numerical results of array syntheses are too rough to capture the coupling effects.

\begin{figure}[htbp]
	\vspace{-0.05in}
	\centerline{\includegraphics[width=3.2in]{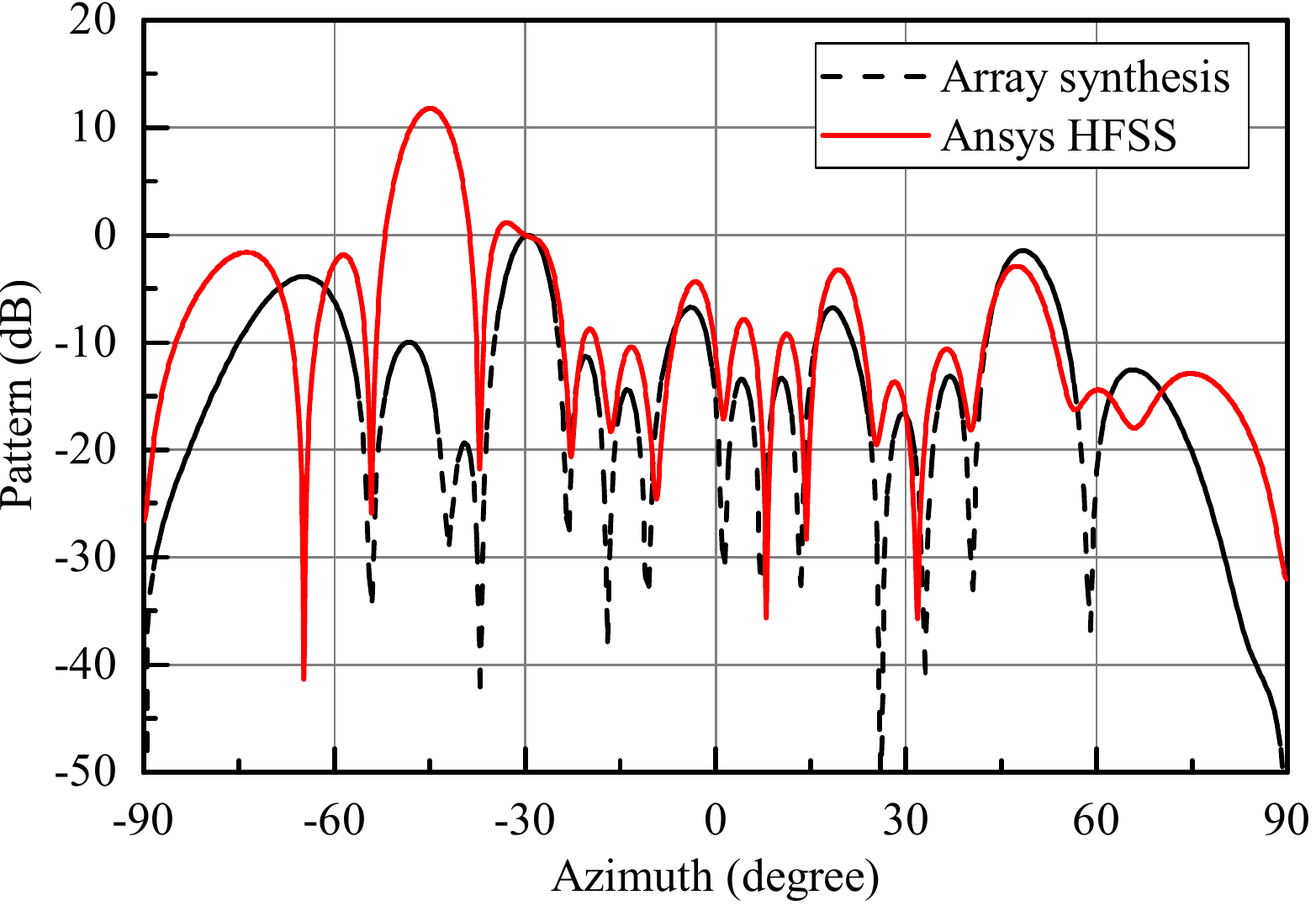}}
	\vspace{-0.05in}
	\caption{Comparison of the pattern between the array synthesis and the full-wave simulation from Ansys HFSS for the $16\times{}1$ DCMs.}
	\label{fig:1}
	\vspace{-0.05in}	
\end{figure}

Let us consider the small-scale $16\times{}1$ 1-bit DCMs along $x$ axis.
The array is synthesized to reflect the TE wave from $\text{azimuth}=\ang{45}$ to $\text{azimuth}=\ang{-30}$ in the $xoy$ plane.
As shown in Fig. \ref{fig:1}, the synthesized reflective pattern concerning the array factors are compared with the full-wave simulation from Ansys HFSS \cite{HFSS19}, in which the coding units in \cite{2020_TAP_DCM} are adopted.
The direction of the main lobe agrees well with each other at $\ang{-30}$,
whereas the overall synthesized results significantly deviates from the full-wave simulations.
In other words, 
the synthesized pattern of DCMs cannot reflect the actual performance without the fully consideration of the coding units.
Besides, some crucial parameters, including the half power beam width and cross polarization discriminant, are invalid, and the cross-polarized gain is also missing in the context.
Therefore, the full-wave simulations are inevitable to validate the final design.

Without loss of generality, the surface integral equations are taken to analyze this problem.
The distributions of the surface currents of the conductors and dielectrics 
are described by the electric field integral equation (EFIE) and the Poggio-Miller-Chang-Harrington-Wu-Tsai (PMCHWT) equations, respectively, 
yielding accurate solutions with the contact-region modeling method \cite{2003_TAP_ADM}.
The equivalent surface currents are then expanded with the Rao-Wilton-Glisson
(RWG) basis functions \cite{TAP_RWG}. 
The method of moments based on the Galerkin discretization leads to the following matrix equation
\begin{align}
	\label{eq:1}
	\textbf{Z}\textbf{I}=\textbf{U},
\end{align}
where \textbf{I} denotes the coefficient vector of the currents, \textbf{U} is the coefficient vector of the excitation and \textbf{Z} is the system matrix.

\subsection{$\mathcal{H}$-matrix Based Fast Algorithm}
Since the DCMs are designed with flexible configurations to fulfill the requested reflective patterns,
a large amount of EM validations are involved.
For the $8\times{}8$ 1-bit DCM, the total number of the possible codebooks is $2^{64}$.
Without regard to periodicity, 
general fast algorithms might have been the best choices to accelerate the full-wave simulation.

\begin{figure}[htbp]
	\centering
	\subfloat[]{\includegraphics[width=1.9in]{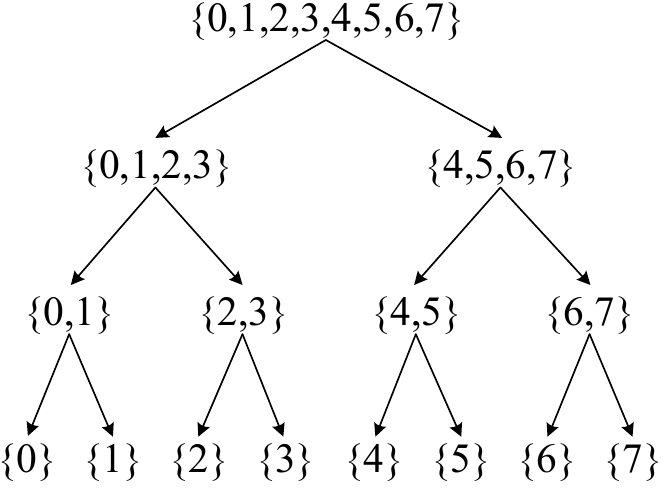}\label{fig:1_p:a}}
	\hspace{0.05in}
	\subfloat[]{\includegraphics[width=1.4in]{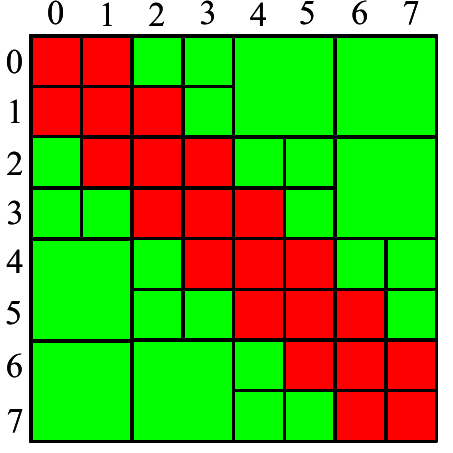}\label{fig:1_p:b}}
	\vspace{0.0in}	
	\caption{An example of the construction of $\mathcal{H}$-matrix. (a) Partition of the cluster tree. (b) Layout of the $\mathcal{H}$-matrix.}
	\label{fig:1_post}
	\vspace{-0.05in}
\end{figure}

The solution to IEs can be accelerated by $\mathcal{H}$-matrix.
In the framework of $\mathcal{H}$-matrix \cite{Borm_Hmatrix_Notebook_2003}, 
the index set of the bases are first partitioned into clusters.
Denote the index set as $\mathcal{I}=\{0,1,\cdots,N-1\}$, where $N$ is the number of the bases. 
The indices in $\mathcal{I}$ are clustered recursively and bisectionally based on the spatial position of the corresponding bases.
The recursion proceeds and returns as the size of the cluster is less than a predefined constant, which is named as $leafsize$.
The resultant cluster tree is denoted by $\mathcal{T_I}$, an example of which is shown in Fig. \ref{fig:1_post}(a).
All the interactions between clusters in $\mathcal{T_I}$ 
are also characterized by a tree,
which is called the block cluster tree and denoted by $\mathcal{T_{I\times{}I}}$.
Assume that clusters $t$ and $s$ lie in the same level of $\mathcal{T_I}$, 
$X_{t}$ and $X_{s}$ are the geometric supports of the bases of $t$ and $s$, respectively.
The block cluster $(t, s)$ is an admissible leaf node in $\mathcal{T_{I\times{}I}}$,
if $t$ and $s$ satisfies the admissibility condition
\begin{align}
	\label{eq:2}
	\max \big\{ \mathrm{diam} (&X_{t}), \mathrm{diam}(X_{s}) \big\}\leqslant{} \eta{}\mathrm{dist}(X_{t}, X_{s}), 
\end{align}
where $\eta$ is a constant coefficient, 
$\rm{diam}(\cdot)$ is the diameter of a cluster,
and $\rm{dist}(\cdot, \cdot)$ is the distance between any two clusters.
Likewise, 
if the admissibility condition is not fulfilled and both $t$ and $s$ are leaf clusters in $\mathcal{T_I}$,
then $(t, s)$ is an inadmissible leaf node in $\mathcal{T_{I\times{}I}}$.
Otherwise, $(t, s)$ is a non-leaf node and the admissibility condition should be examined recursively.
Intuitively, (\ref{eq:2}) is the sufficient condition to check whether the distance between two clusters is relatively far enough and the interaction is weak.
If $(t, s)$ is admissible, the corresponding matrix $\bm{Z}_{t, s}$ allows low-rank approximation with purely algebraic algorithms, 
such as ACA \cite{Bebendorf_ACA}.
An example of the $\mathcal{H}$-matrix corresponding to Fig. \ref{fig:1_post}(a) is illustrated in Fig. \ref{fig:1_post}(b),
where the admissible blocks are colored in green, and inadmissible ones are in red.
The C code for the construction of $\mathcal{H}$-matrix are presented in Chapter 2 of \cite{Borm_Hmatrix_Notebook_2003}. 

\section{Methodology for Further Acceleration}
\label{sec:methodology}

EM validations of DCMs with varied codebooks involves the simulations of akin structures, 
which makes it special in several aspects.
First of all, the codebook of one DCM shares a large portion of the coding states with some other ones.
Intuitively, it is unwise to treat theses configurations separately and execute validations from scratch with the presence of solutions to similar ones.
Additionally, the total number of possible codebooks of the DCMs is still limited, 
which essentially differs from the parameterized topology with continuous variables.
Furthermore, the overall structures of DCMs are quasi-periodic, which means the system matrix highly resembles that of the rigorously periodic arrays.
Intuitively, it is desirable to fully exploit all the above features to achieve more significant acceleration than general fast algorithms.
To this end, we propose to form rational periodization,
which has the the potential to further improve the efficiency. 

\subsection{Recurrence Rebuild and Retrieval Method}
Consider the $k$-bit DCMs with an array size of $m\times{}n$. 
Despite the $2^{k}$ coding states, the collection of the set of all candidates is unique.
The geometrical union of the collection is taken to compose the macro unit.
Once each coding unit is replaced by the macro unit, 
the overall structure is treated as rigorously periodic.
The formation of periodicity from general arrays via the substitution with macro units is named as the recurrence rebuild.

\begin{figure}[htbp]
	\centering
	\vspace{-0.0in}
	\subfloat[]{\includegraphics[width=1.5in]{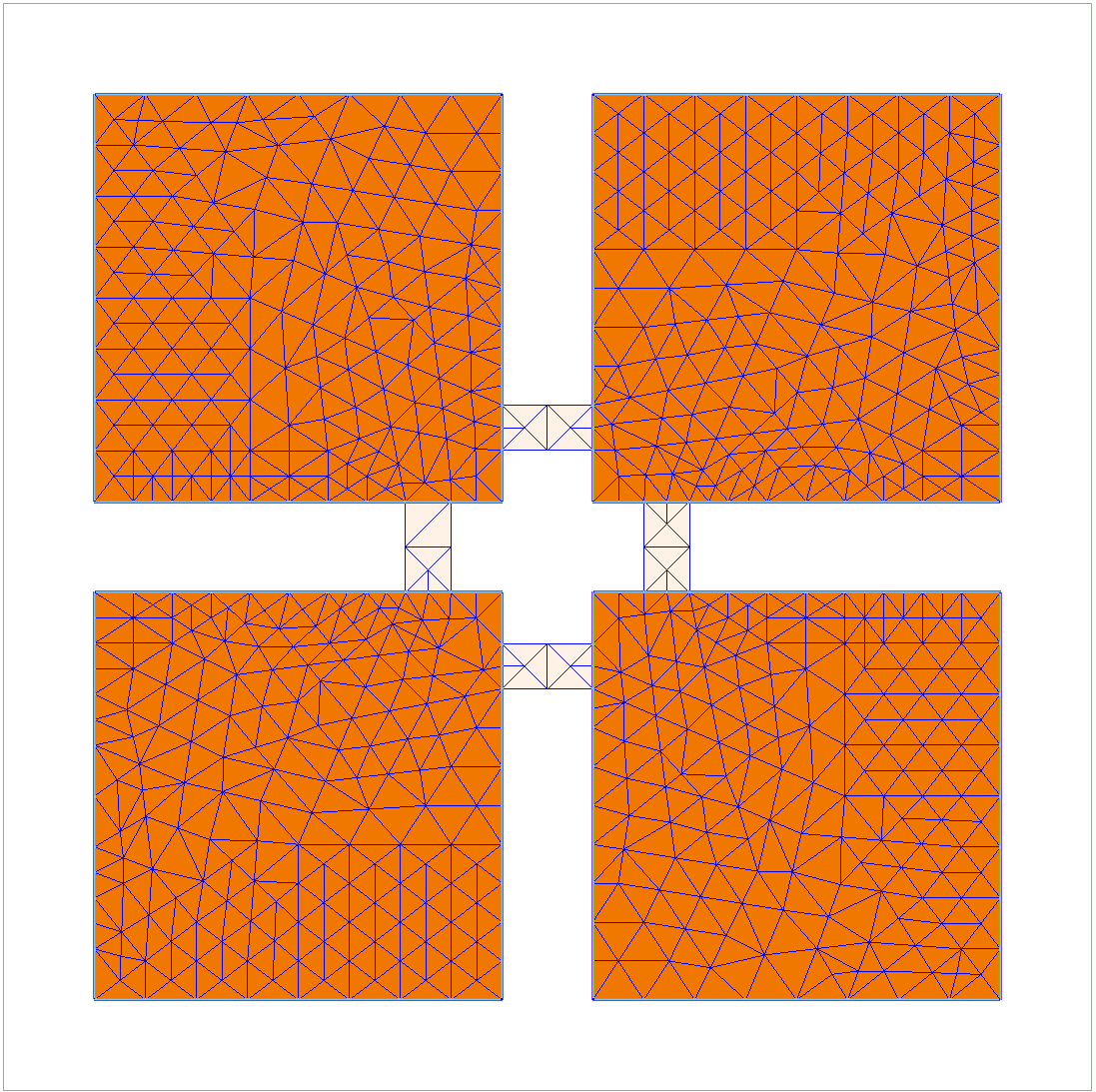}\label{fig:3:a}}
	\hspace{0.15in}
	\subfloat[]{\includegraphics[width=1.5in]{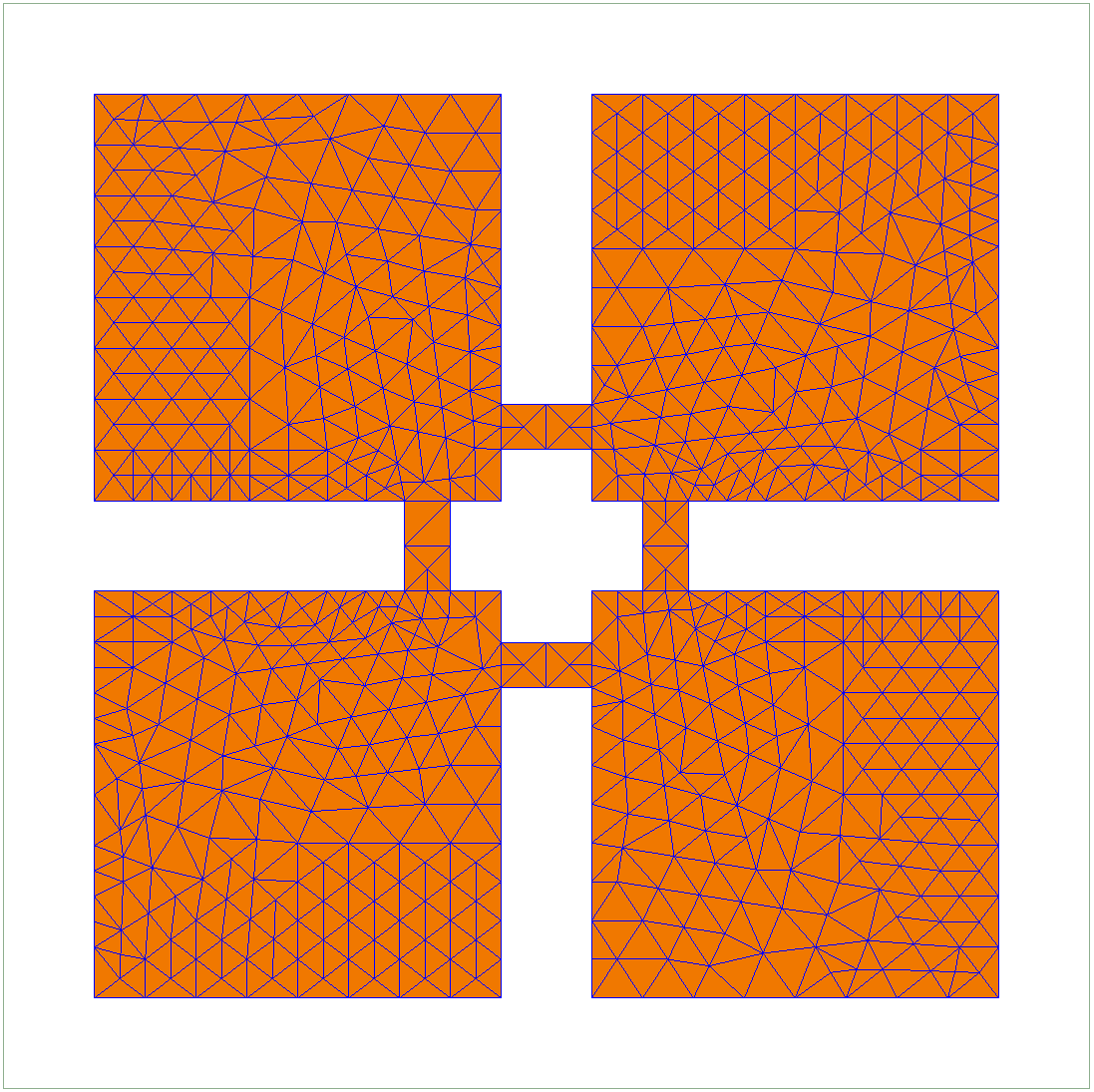}\label{fig:3:b}}\\
	\vspace{-0in}
	\subfloat[]{\includegraphics[width=1.5in]{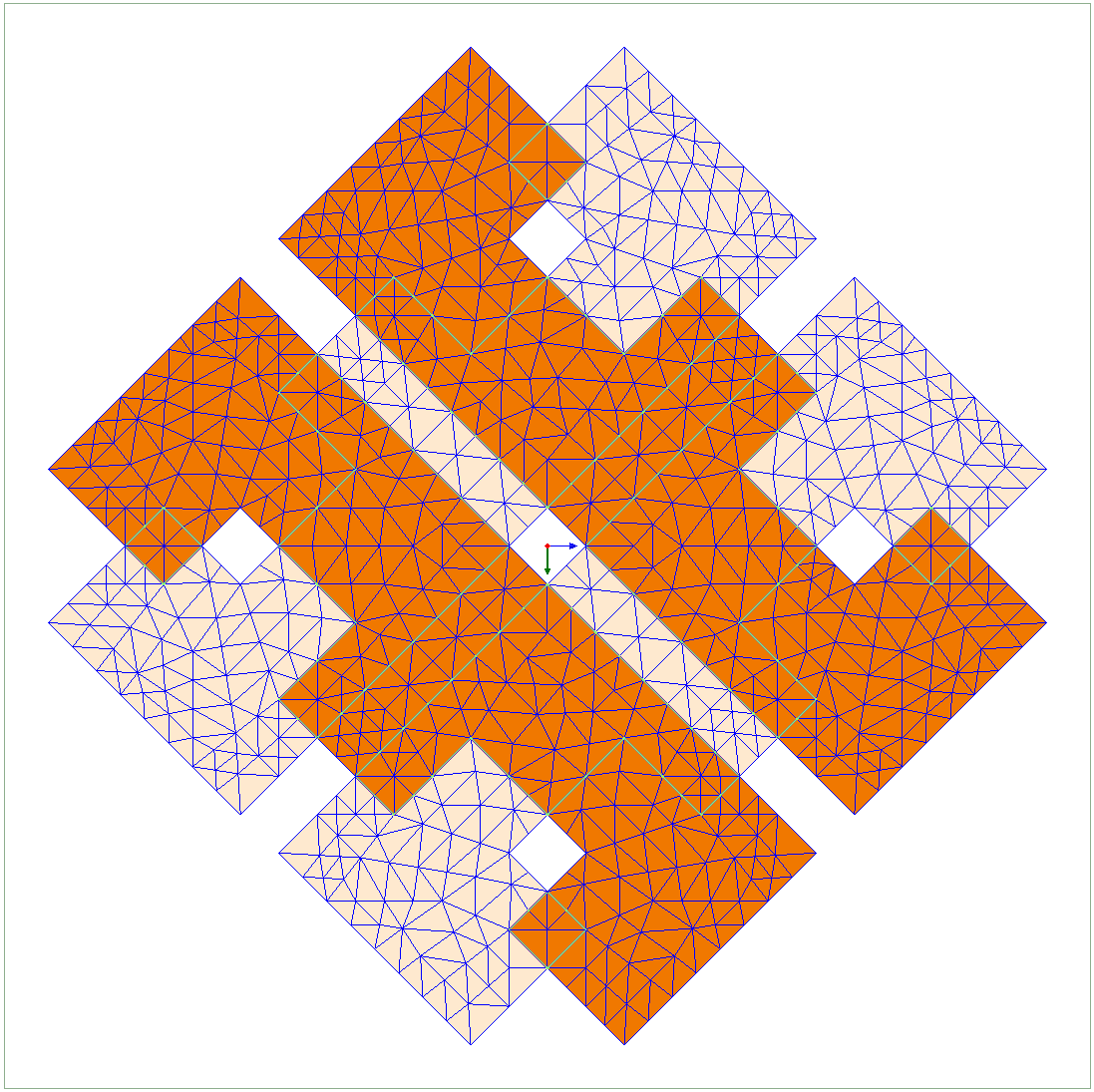}\label{fig:3:c}}
	\hspace{0.15in}
	\subfloat[]{\includegraphics[width=1.5in]{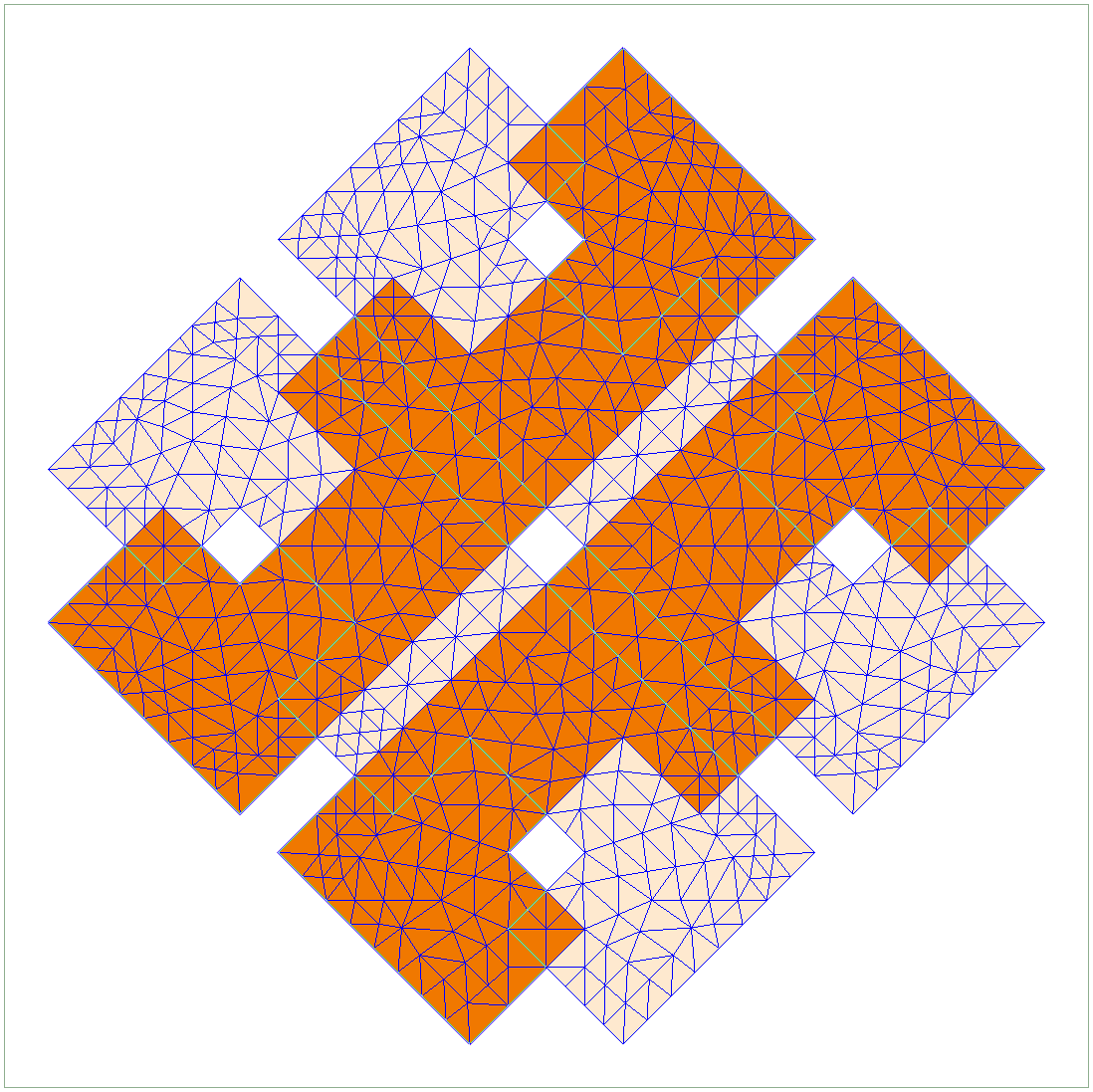}\label{fig:3:d}}\\		
	\caption{Illustrations of two types of 1-bit coding units and macro units. 
		(a) State 0 of the type-A unit. 
		(b) State 1 of the type-A unit. 
		(c) State 0 of the type-B unit. 
		(d) State 1 of the type-B unit. 
	}
	\label{fig:3}
	\vspace{-0.in}
\end{figure}

With recurrence rebuild, the $2^{k}$ coding states are mixed-up together shaping the macro unit,
which is discretized as a whole and leads to coexistent meshes.
The commonly used coding units are of $1$-bit or $2$-bit, meaning $2$ or $4$ coding states.
As shown in Fig. \ref{fig:3}, 
the structures highlighted in Fig. \ref{fig:3}(a) and Fig. \ref{fig:3}(b) compose a pair of 1-bit states of the connectivity-controlled coding units, which was proposed in \cite{2020_TAP_DCM} and denoted by type-A.
The structures highlighted in Fig. \ref{fig:3}(c) and Fig. \ref{fig:3}(d) compose another pair of 1-bit states of the rotation-controlled coding units, which was proposed in \cite{2018_ACCESS} and denoted by type-B.
Each subfigure illustrates the corresponding macro unit, 
in which the difference set between the union and the coding unit is virtually presented in light color. 
Each unit thereby constitutes a subset of the macro unit. 

The structure of the array of macro units is denoted by $\mathcal{S}$, 
and the structure of one instance of DCMs is denoted by $\mathcal{S}_{k}$.
Apparently, $\mathcal{S}_{k}$ constitutes a subset of $\mathcal{S}$.
Assume that $\mathcal{S}$ is discretized by $N$ bases $\{\bm{f}_{i}(\bm{r})\}$, $i=0,1,\cdots,N-1$, and $\mathcal{S}_{k}$ occupies some $N_{k}$ bases ($N_{k}\leq{}N$).
Note that the matrix equation for $\mathcal{S}_{k}$ has the identical form of (\ref{eq:1}).
The system matrix $\textbf{Z}_{k}$ for $S_{k}$ should be exactly the retrieval of the matrix $\textbf{Z}$ that corresponds to the system matrix discretizing $\mathcal{S}$,
likewise the excitation vector.
It is equivalent to set unrelated entries in $\textbf{Z}$ and $\textbf{U}$ to zeros 
and solve only part of $\textbf{I}$.

\begin{figure}[htbp]
	\centering
	\includegraphics[width=3.4in]{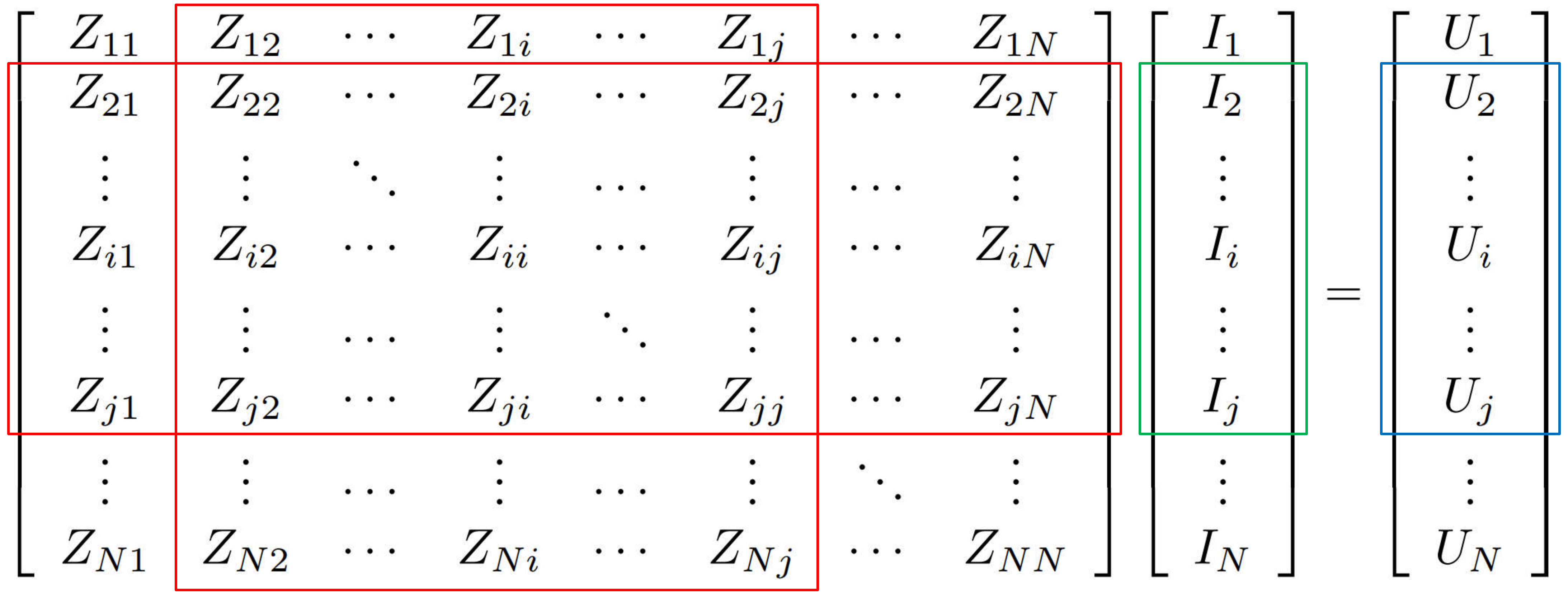}
	\caption{Explicit retrieval process for the construction of matrix equation.}
	\label{fig:4}
\end{figure}

The retrieval is executed based on whether the spatial region of each basis function belongs to $\mathcal{S}_{k}$.
$\textbf{Z}$ is shared across the solving process for all codebooks and named as the global matrix.
For example, $\mathcal{S}_{k}$ is assumed to occupy the basis functions labeled from $2$ to $j$.
The Matlab's slicing notations for the system matrix and excitation vector of $\mathcal{S}_{k}$ are $\textbf{Z}_{k}=\textbf{Z}(2\colon{}j,2\colon{}j)$
and $\textbf{U}_{k}=\textbf{U}(2\colon{}j)$.
This explicit retrieval example is visualized in Fig. \ref{fig:4}.
In spite of generality, the explicit retrieval consumes nearly twice the storage during computation.
Moreover, prevailing fast algorithms, including FMM and $\mathcal{H}$-matrix, do not have explicit expressions for far-field interactions, making 
the explicit process inadvisable. 

Nevertheless, the implicit matrix retrieval implemented via the matrix-matrix product still provide efficient solutions.
The matrix slicing the corresponding columns of $\textbf{Z}$ for $\mathcal{S}_{k}$ is denoted by $\textbf{P}_{k}$, 
of which the $j$-th column is the standard unit vector with the ${n_{j}}$-th element being $1$ if 
$S_{k}$ occupies $\bm{f}_{n_{j}}$.
Obviously, $\textbf{P}_{k}$ is a rectangular matrix of shape $N\times{}N_{k}$.
The implicit retrieval of $\textbf{Z}_{k}$ from the global matrix $\textbf{Z}$ for $\mathcal{S}_{k}$ gives
\begin{equation}
\textbf{Z}_{k} = \textbf{P}_{k}^{\text{T}}\textbf{Z}\textbf{P}_{k},
\label{eq:3}
\end{equation}
which is equivalent to Matlab's slicing notations.
The coefficients in $\textbf{U}$ and $\textbf{I}$ that correspond to the basis functions that do not belong to $\mathcal{S}_{k}$ are fixed as zeros, and we have
\begin{equation}
	\begin{aligned}
		&\textbf{I} = \textbf{P}_{k}\textbf{I}_{k}, \;\;
		&\textbf{U} = \textbf{P}_{k}\textbf{U}_{k},\\
		&\textbf{I}_{k} = \textbf{P}_{k}^{\text{T}}\textbf{I}, \;\;
		&\textbf{U}_{k} = \textbf{P}_{k}^{\text{T}}\textbf{U}.
	\end{aligned}
	\label{eq:4}
\end{equation}
According to the constraints in (\ref{eq:3}) and (\ref{eq:4}),
the system matrix equation for each specific $\mathcal{S}_{k}$ is then reformulated as
\begin{equation}
\textbf{P}_{k}\textbf{P}_{k}^{\text{T}}\textbf{Z}\textbf{P}_{k}\textbf{P}_{k}^{\text{T}}\textbf{I}=\textbf{P}_{k}\textbf{P}_{k}^{\text{T}}\textbf{U}.
\label{eq:5}
\end{equation}

Considering the definition of $\textbf{P}_{k}$, the matrix $\textbf{P}_{k}\textbf{P}_{k}^{\text{T}}$ is strictly diagonal, with the entries being $1$'s or $0$'s.
Without explicitly forming $\textbf{Z}_{k}$ or constructing the slicing matrix $\textbf{P}_{k}$, the equation (\ref{eq:5}) is further simplified as
\begin{equation}
\textbf{D}_{k}\odot{}\left(\textbf{Z}\left(\textbf{D}_{k}\odot{}\textbf{I}\right)\right)=\textbf{D}_{k}\odot{}\textbf{U},
\label{eq:6}
\end{equation}
where $\textbf{D}_{k}=\text{diag}(\textbf{P}_{k}\textbf{P}_{k}^{\text{T}})$ is the flattened indexing vector of the diagonal of $\textbf{P}_{k}\textbf{P}_{k}^{\text{T}}$, and $\odot$ denotes the Hadamard product.

We call the above periodization procedures for DCMs as the recurrence rebuild and retrieval method (R3M).
Note that R3M 
suffices well the requirements for fast EM validations.
The global matrix is able to be calculated once in the beginning and shared across the solving process,
during which the matrix slicing via implicit retrieval are capable to adjust to arbitrary codebooks. 
In addition, recurrence rebuild fully leverages the special quasi-periodicity of DCMs,
enabling more efficient and compact representations of the global matrix $\textbf{Z}$.
The matrix-vector product (MVP) based iterative methods support efficient solutions,
whereas the direct solutions to linear equations, 
including the LU decomposition and inversion, 
can hardly inherit information from $\textbf{Z}$ due to the variability of $\textbf{D}_{k}$.

Note that R3M will increase the number of unknowns,
especially when the coding states have widely different meshing densities, 
although it is not a general case.
The periodization via R3M is still worthwhile when the number of unknowns increases within several times,
since a compact form of the global matrix is close at hand together with HPE \cite{2021_TAP_HPE}.


\subsection{Basis Extension of Periodic Boundaries}
\begin{figure}[htbp]
	\centering
	\subfloat[]{\includegraphics[width=2.2in]{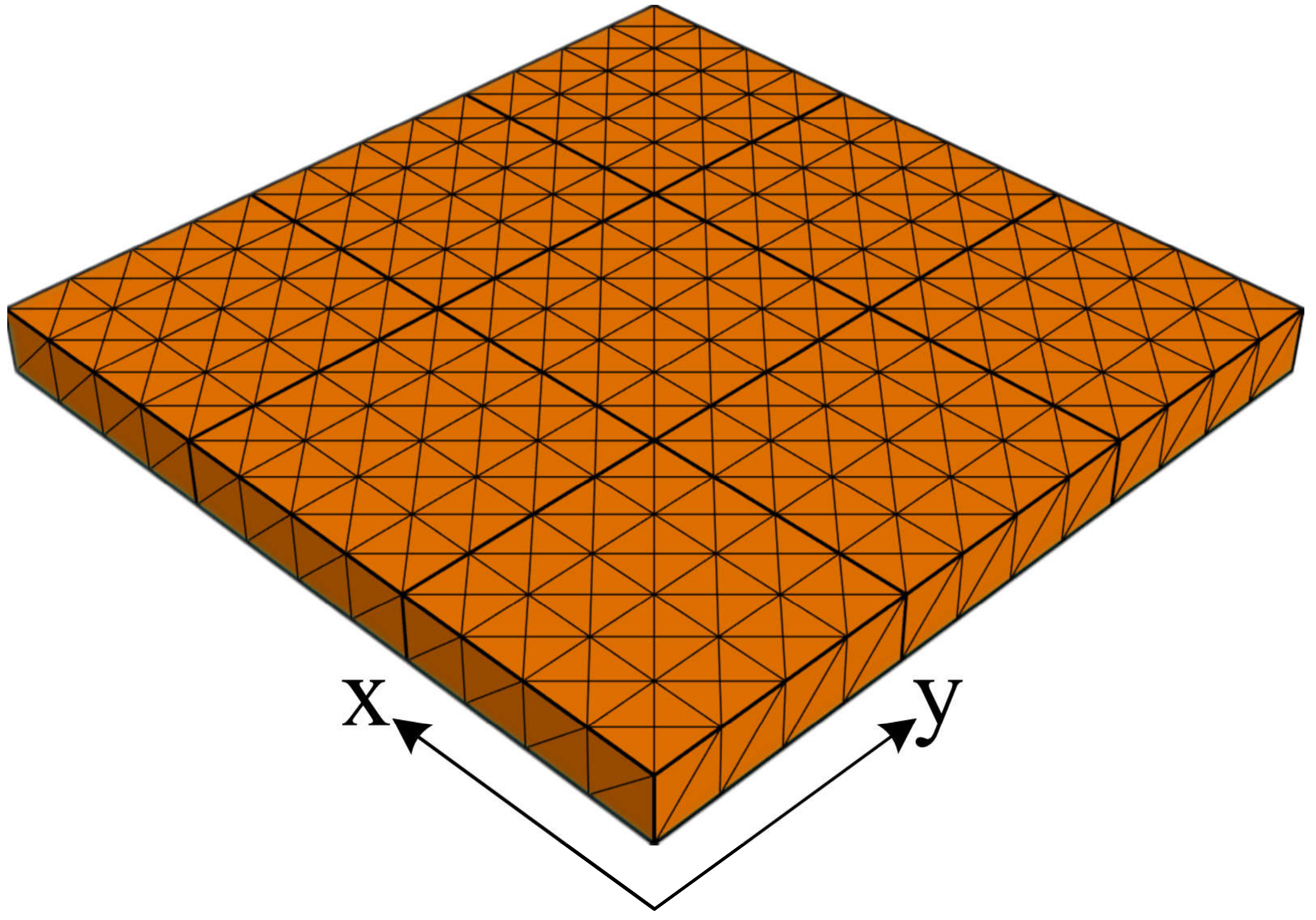}\label{fig:5:a}}\\	
	\vspace{0.0in}
	\subfloat[]{\includegraphics[width=2.2in]{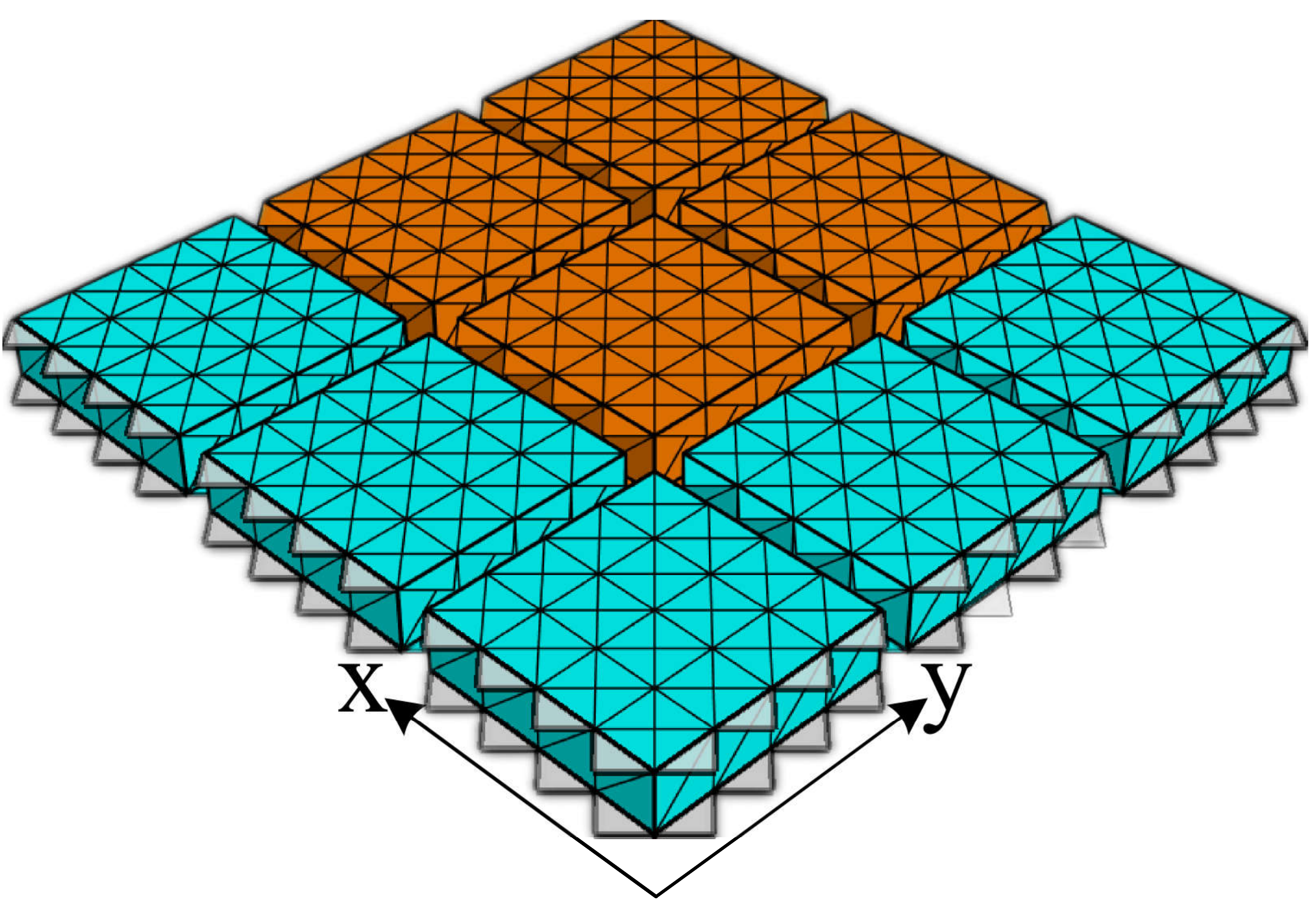}\label{fig:5:b}}
	\vspace{0.0in}	
	\caption{Basis extension of periodic boundaries for the substrate. (a) Original physical model. (b) Equivalent $3\times{}3$ slab array model.}
	\label{fig:5}
	\vspace{-0.0in}
\end{figure}

Conventionally,
$8$ types of edge and corner units are specially handled, 
at the sacrifice of effective exploitation of the periodicity.
Since the RWG basis function is composed of two triangular facets with common edge \cite{TAP_RWG},
the basis that bridge over adjacent coding units should be ascribed to either one, 
for the sake of the continuity of the edge current along periodic boundaries.
The bases at the boundaries, which only exist in the edge units,
are extended to ensure rigid periodicity in the mesh.
The ground plane is modeled as a connected patch array, and
the substrate is modeled as a contacted slabs array, as showcased by the contacted $3\times{}3$ slab array in Fig. \ref{fig:5}(a).

The basis extension leads to deviations from the physical model,
due to the presence of the extended bases.
Fortunately, it is convenient to adopt the retrieval methods to resolve the problem by appropriately setting indexing vector $\textbf{D}_{k}$ in (\ref{eq:6}).
As shown in Fig. \ref{fig:5}(b), the edge units colored in blue
should have extended bases towards $-x$ and/or $-y$ axes to keep the rigid periodicity in the mesh.
For a clearer exposition, 
the extended bases are virtually presented by setting transparency and filled with white color.
The entries in $\textbf{D}_{k}$ that correspond to the extended bases are masked with zeros to accord with the physical model, which fits well in the framework of R3M.

\subsection{Hierarchical Pattern Exploitation for Periodic Array}

With the coding units being replaced by the macro units,
the DCM is transformed to periodic array that enables concise representation.
The periodic array is uniquely determined by the layout,
which is jointly expressed by the periodic lattice and scales.
It is expected to exploit the periodicity and spatial information simultaneously
to refrain from repeated calculations.
Since $\mathcal{H}$-matrix treats the bases separately,
hierarchical clustering of the indices 
takes into account
only the spatial information, 
and overlooks their assignments to specific units. 
For that reason, 
we propose to adopt the hierarchical pattern exploitation (HPE) by default 
to cooperate with R3M, denoted by R3M-HPE, for fully exploitation of the rigid periodicity.

HPE was first proposed in \cite{2021_TAP_HPE}.
We describe the algorithm briefly here.
The core idea of HPE is inspired by the observation that a periodic array is capable to be recursively represented without the loss of computational feasibility.
This idea provides essential insights for the efficient representation of the system matrices of periodic arrays.
Following a similar line of thought,
the spatial partitioning is performed over the periodic lattice, 
which leads to the object cluster tree denoted by $\mathcal{T_{O}}$.
The object cluster tree is also built bisectionally and recursively.
The recursion returns if the object cluster contains only one array element, 
which is the finest object cluster as leaf node.
The object cluster tree retains the structural feature of the array without diving into the interior spatial information of each element.
The interaction between object clusters in $\mathcal{T_{O}}$ is characterized by the block object tree
$\mathcal{T_{O\times{}O}}$.
The nodes in $\mathcal{T_{O\times{}O}}$ representing interactions between object clusters are denoted as patterns and will be defined shortly.
Obviously,
the finest operating unit in $\mathcal{T_{O\times{}O}}$ is the $1\times{}1$ block, 
representing the interaction between two array elements.
The bases within each unit are further bisectionally partitioned, 
following the leaf node of the object cluster tree,
to obtain the final cluster tree.
The admissibility condition (\ref{eq:3}) is examined during the construction of the block cluster tree the same as $\mathcal{H}$-matrix.

\begin{figure}[htbp]
	\centering
	\vspace{-0.in}
	\subfloat[]{\includegraphics[width=1.5in]{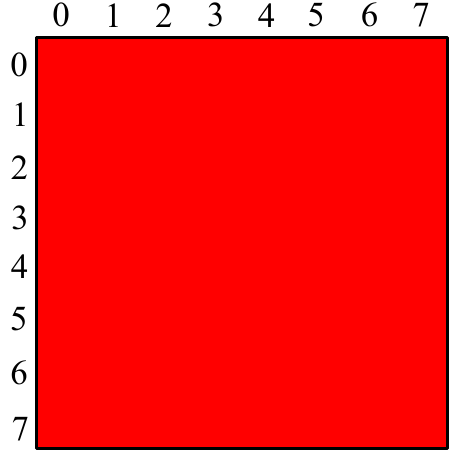}\label{fig:6:a}}
	\hspace{0.15in}
	\subfloat[]{\includegraphics[width=1.5in]{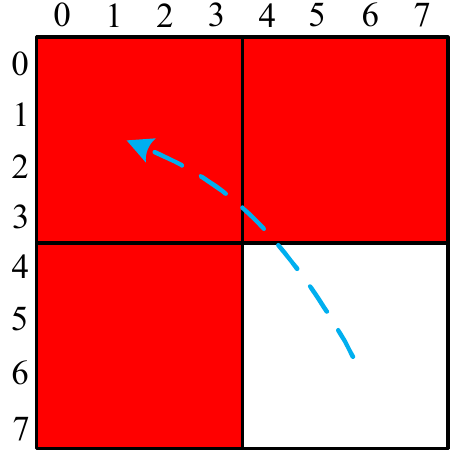}\label{fig:6:b}}\\
	\subfloat[]{\includegraphics[width=1.5in]{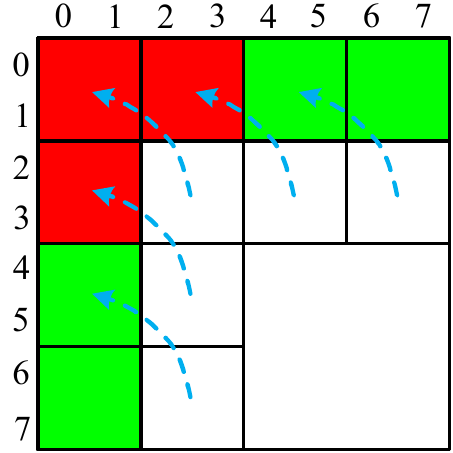}\label{fig:6:c}}
	\hspace{0.15in}
	\subfloat[]{\includegraphics[width=1.5in]{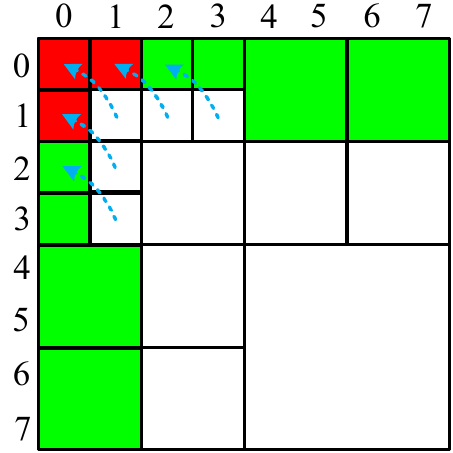}\label{fig:6:d}}
	\caption{Layouts of the virtual $\mathcal{H}$-matrix at different levels with HPE for the $1\times{}8$ periodic array. 
	(a) 1st level. (b) 2nd level with $1$ shared pattern. (c) 3rd level with $5$ shared patterns. (d) 4th level with $5$ shared patterns.}
	\label{fig:6}
	\vspace{-0in}
\end{figure}

For periodic arrays, 
the object clusters within $\mathcal{T_{O}}$ are jointly
depicted by the shapes and starting coordinates.
However,
the absolute starting coordinates of the source and observer are superfluous for the block object clusters, and only the relative coordinate matters.
Therefore, the pattern can be characterized by the tuple
\begin{equation}
\text{Pattern}:=\left(\text{offset}, \text{src}, \text{obs}\right),
\label{eq:7}
\end{equation}
where the $\text{offset}$ denotes the relative offset coordinate from the coordinate of the source to that of the observer, the $\text{src}$ and $\text{obs}$ denote the shapes of the source and observer object clusters, respectively.

HPE represents the reuse of patterns at multi-levels within the hierarchy of $\mathcal{H}$-matrix.
Take the $1$-D periodic array with $8$ elements labeled sequentially from $0$ to $7$ as an example.
Without diving into the interior of each element,
the cluster tree and the corresponding $\mathcal{H}$-matrix are exactly the same as those shown in Fig. \ref{fig:1_post}.
The application of HPE in this scenario
with $\mathcal{H}$-matrix 
are illustrated level by level in Fig. \ref{fig:6}.
The $1$th level of $\mathcal{H}$-matrix is full-rank without partition.
The matrix is then divided into $4$ submatrices at the $2$nd level with one reused pattern,
due to the fact that the blocks $(0,\cdots,3)\times{}(0,\cdots,3)$ and $(4,\cdots,7)\times{}(4,\cdots,7)$ share the same pattern.
The reuses of patterns are visualized by the blue dashed lines with arrows.
At the $3$rd level, the number of reused patterns is $5$, 
likewise the $4$th level, 
as shown in Fig. \ref{fig:6}(c) and \ref{fig:6}(d), respectively.
The white blanks in Fig. \ref{fig:6} indicate that they reference other subblocks with identical patterns,
which are shared across the $\mathcal{H}$-matrix.
The incomplete $\mathcal{H}$-matrix obtained from HPE with shared subblocks is denoted as the virtual $\mathcal{H}$-matrix.
HPE is implemented by hashing the pattern and creating a dictionary 
that maps patterns to the corresponding submatrices. 
Hence,
comparing with the layout of $\mathcal{H}$-matrix in Fig. \ref{fig:1_post}(b),
a large amount of redundant computations in Fig. \ref{fig:6}(d) are pruned with HPE,
which is theoretically more efficient than the classical $\mathcal{H}$-matrix for the matrix assembly of periodic arrays.



\subsection{Near-Field $\mathcal{H}$-LU Preconditioner}
The equation (\ref{eq:6}) can be solved with iterative methods, even though the matrix $\textbf{Z}$ is represented 
in the form of
virtual $\mathcal{H}$-matrix.
A proper preconditioner is necessary to obtain rapid convergence.
Within the framework of $\mathcal{H}$-matrix, the $\mathcal{H}$-LU can obtain a efficient preconditioner \cite{2019_AWPL_LU}.
However, direct LU decomposition is relatively time-consuming and wasteful for iterative solutions.
In contrast, the near-field preconditioner has been heavily studied and is robust and efficient \cite{2017_SIAM}.

The near-field interactions are represented by full-rank submatrices in $\mathcal{H}$-matrix.
The ignorance of the far-field interactions equals to set the low-rank submatrices to be all zeros.
From this perspective, the $\mathcal{H}$-LU composition of the near-field is equivalent to the block incomplete LU decomposition \cite{2003_Iterative}.
It should be noted that the inversion of the retrieval of the matrix differs from the retrieval of the inversion, 
and so does the $\mathcal{H}$-LU decomposition.
The preconditioner for each specific codebook supports efficient construction via the retrieval from the full-rank submatrices within the global matrix. 

\subsection{Analysis of Computational Complexities}
For the classical $\mathcal{H}$-matrix, the storage cost of the low-rank matrices at the $l$-th level scales as
\begin{equation}
	\label{eq:8}
	S_l = C_{sp}2^{l}\times{}2k\frac{N}{2^{l}}=2kC_{sp}N,
\end{equation}
where $C_{sp}$ is a constant denoting the maximal number of blocks formed by every single cluster,
$k$ is the rank of the low-rank matrices, and $N$ is the number of unknowns \cite{Borm_Hmatrix_Notebook_2003, 2013_CPMT_Chai}.
For the virtual $\mathcal{H}$-matrix resulted from HPE, 
the storage of the low-rank matrices at the $l$-th level is evaluated by
\begin{equation}
	\label{eq:9}
	S_{l}^{\prime} = C_{sp}\times{}2k\frac{N}{2^{l}}=\frac{S_l}{2^{l}},
\end{equation}
where the $2^l$ representing the number of clusters is 
eliminated in contrast with (\ref{eq:8}), leading to $2^l$ times storage reduction at the $l$-th level.
The total amount of storage is given by
\begin{equation}
	\label{eq:10}
	S^{\prime}=\sum_{l=1}^{L}S_{l}^{\prime} = \mathcal{O}(kN).
\end{equation}

Similarly, 
in terms of the CPU time cost,
the ACA based low-rank matrix assembly is also $2^l$ times faster than the classical $\mathcal{H}$-matrix
at the $l$-th level, 
except that the coefficient related to the rank is replaced by the squared rank \cite{Bebendorf_ACA} as
\begin{equation}
	\label{eq:11}
	T^{\prime}=\sum_{l=1}^{L} C_{sp}\times{}k^2\frac{N}{2^{l}} = \mathcal{O}(k^{2}N).
\end{equation}

The number of submatrices representing near-field interactions is shrunk to constant with the fully reuse of patterns.
In the scenario of DCMs, the geometrical structure is stretched in the 2-D plane.
Thus, the maximum rank $k$ increases relatively slowly as $\mathcal{O}(\sqrt{\log{N}})$ with respect to the number of basis functions \cite{2013_CMPT_RANK}.
Moreover, the computational complexity of $\mathcal{H}$-LU is bounded by $\mathcal{O}(N\log{N})$ for the near-field $\mathcal{H}$-matrix.
Though the MVP is of a higher complexity as $\mathcal{O}(kN\log{N})$,
the final complexity is bounded by $\mathcal{O}(N\log{N})$, due to the fast convergence of the iterative solution.


\section{Numerical Examples}
\label{sec:examples}
\begin{figure}[!b]
	\centering
	\vspace{-0.2in}
	\subfloat[]{\includegraphics[width=1.65in]{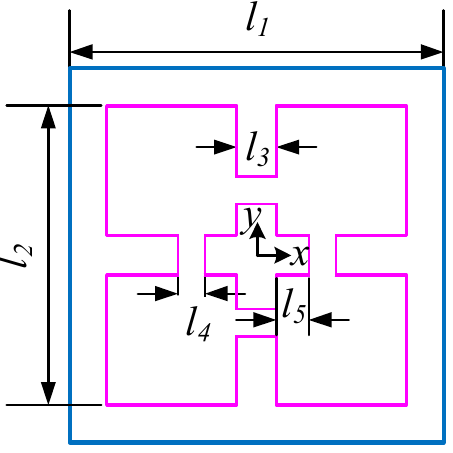}\label{fig:b:a}}
	\hspace{0.05in}
	\subfloat[]{\includegraphics[width=1.65in]{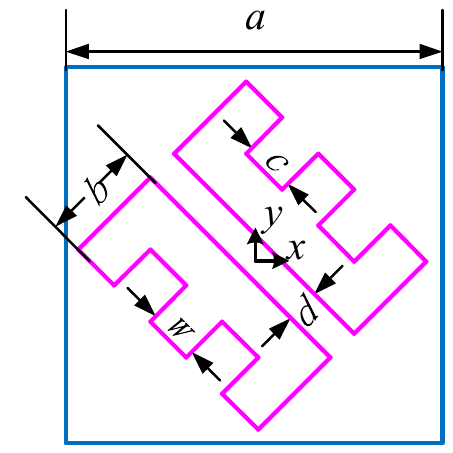}\label{fig:b:b}}
	\caption{Top view of the 1 bit unit cell of the metasurface. (a) State $1$ of type-A unit, where $l_1=3.60$, $l_2=3.00$, $l_3=0.30$, $l_4=0.15$, and $l_5=0.17$. (b) State $0$ of type B unit, where $a=2.0$, $b=0.5$, $c=0.3$, $d=0.1$, and $w=0.3$ (unit:mm).}
	\label{fig:7}
	\vspace{-0in}
\end{figure}

Two types of 1-bit DCMs are studied in this section to check the performance of the proposed method in electromagnetic validations, as shown in Fig. \ref{fig:7}.
BiCGStab \cite{2003_Iterative} is adopted as the iterative solver with a tolerance of $10^{-6}$ for all cases in this section. 
The computing platform is the desktop PC with Intel Core I9-9900K CPU at $3.60$ GHz and $64$ GB RAM. 
Although $\mathcal{H}$-matrix supports efficient parallelization of matrix operations \cite{2019_Parallelization},
the single-thread mode is utilized to count the CPU time for fair comparisons.
The accuracy, convergence and efficiency of R3M-HPE are examined to validate its feasibility.
Hereinafter, all directional angles are presented in azimuth and elevation angles of radar coordinate, appearing as $(\text{azimuth}, \text{elevation})$.
Both the classical $\mathcal{H}$-matrix and the MLFMA is used to make comparisons.
The classical $\mathcal{H}$-matrix is implemented by our in-house code, and the MLFMA in Ansys HFSS is applied.

\subsection{Accuracy}

\subsubsection{Single-Beam Reflective Pattern}
We first inspect the precision of the far-field reflective patterns for the $32\times{}32$ type-A DCM.
The coding units corresponding to State 0 and State 1 are shown in Fig. \ref{fig:3}(a) and Fig. \ref{fig:3}(b), respectively,
and the conductor patches are $1$ mm above the metallic ground.
The DCM is working at $50$ GHz.
The TM wave incidents from $(\ang{45}, \ang{0})$, and the codebook is designed to reflect the wave towards $(\ang{-30}, \ang{0})$.

\begin{figure}[htbp]
	\vspace{-0.05in}
	\centerline{\includegraphics[width=3.1in]{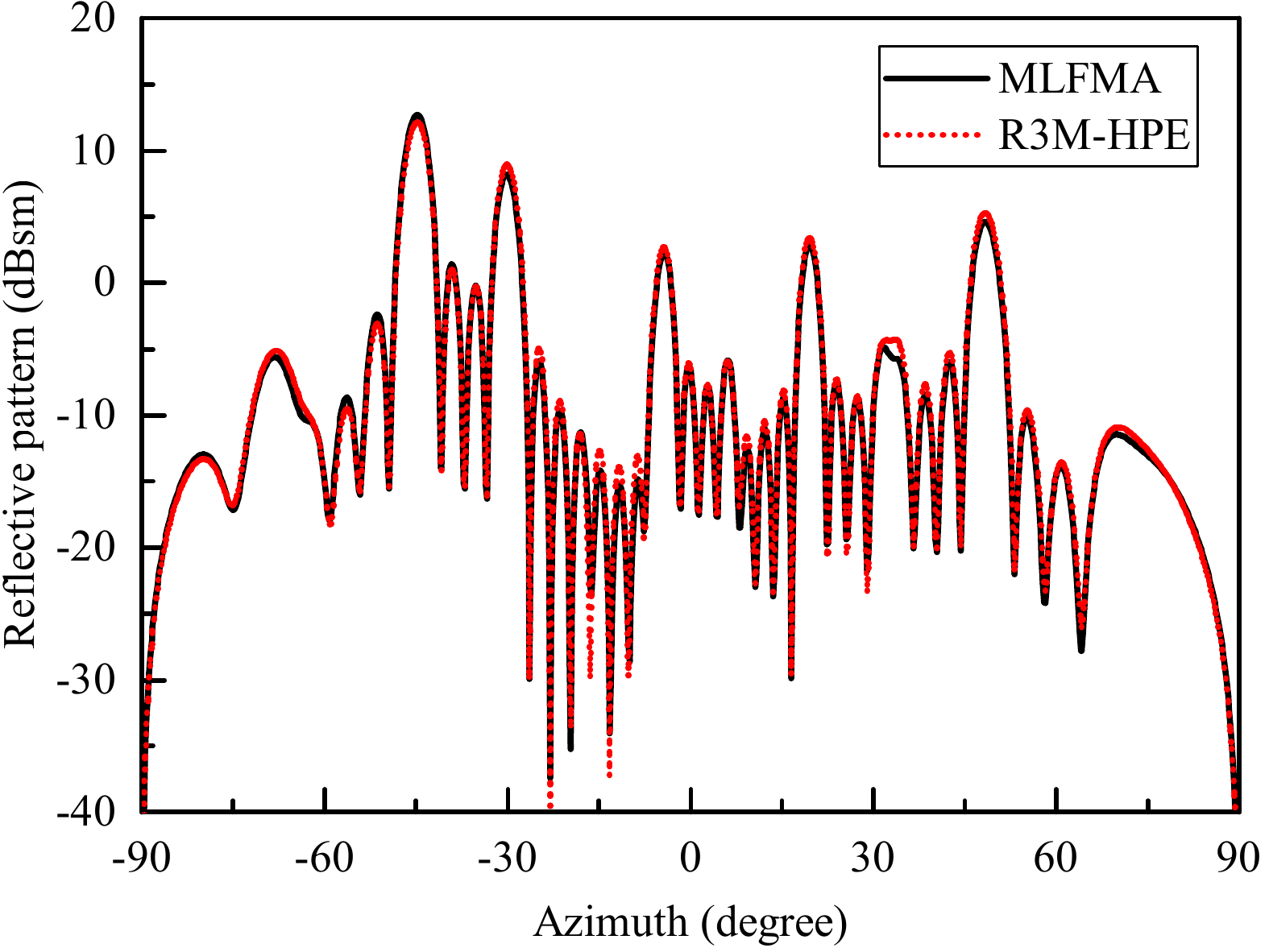}}
	\vspace{-0.10in}
	\caption{Reflective pattern of the $32\times{}32$ type-A DCM.}
	\label{fig:8}
	\vspace{-0.in}	
\end{figure}

Each macro unit of the type-A 1-bit DCM is discretized with $1056$ bases.
Numerical result from R3M-HPE is compared with that from MLFMA.
The reflective pattern in the $(\cdot, \ang{0})$ plane is shown in Fig. \ref{fig:8}.
The beam at $(\ang{-45}, \ang{0})$ is very high due to specular reflection effect.
The overall curves match well with each other, which indicates that the proposed method is accurate with respect to the far-field reflective pattern.

\subsubsection{Surface Current Distribution}
The type-A DCM is still adopted to inspect the electric current distribution at the surface of conductors.
The working frequency and discretization are identical with the former example.
The array size is $8\times{}8$ to facilitate viewing.
The EM wave incidents normally towards $-z$ and polarizes along $x$ axis, and the codebook is designed to reflect the wave towards $(\ang{30}, \ang{0})$.

\begin{figure*}[!t]
	\centering
	\vspace{-0.05in}
	\subfloat[]{\includegraphics[height=2.53in]{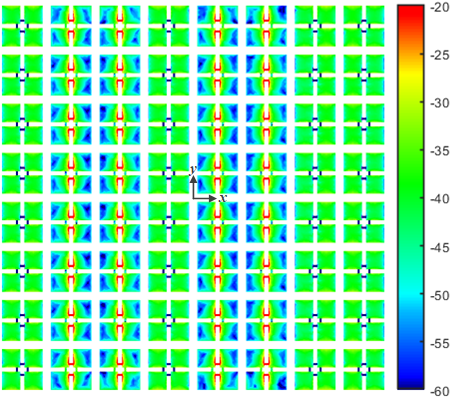}\label{fig:8:a}}
	\hspace{0.5in}
	\subfloat[]{\includegraphics[height=2.54in]{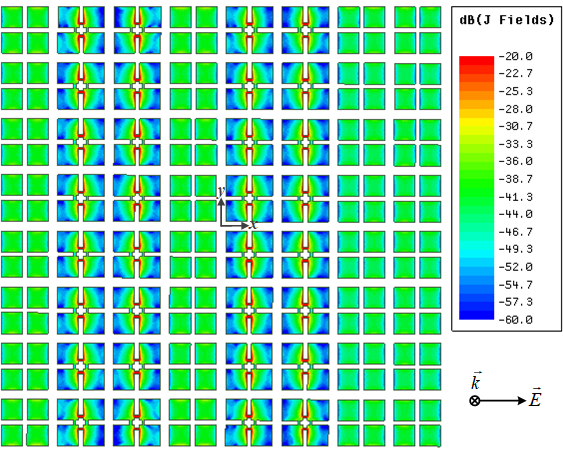}\label{fig:8:b}}\\
	\caption{Surface current distribution of the conductor patches of the $8\times{}8$ type-A 1-bit DCM. (a) Results from R3M. (b) Results from MLFMA. (unit: dBV/m)}
	\vspace{-0.2in}
	\label{fig:9}
\end{figure*}

The current distribution of the conductor patches is compared with the result from MLFMA.
As is observed in Fig. \ref{fig:9},
the current distribution agrees well with each other,
even though the macro units replace the original coding units.
The main difference lies in the fact that the thin strips in between the four patches are removed for State $0$ in the HFSS model, 
while retained in R3M-HPE, 
at the $1$-, $4$-, $7$- and $8$-th columns of the array.
Despite the virtual presence of the thin strips, 
the corresponding coefficients in R3M are forced to be zeros via $\textbf{D}_{k}$ in (\ref{eq:6}),
ensuring the compliance with the physical model.

\subsubsection{Multi-Beam Reflective Pattern}
The type-B 1-bit DCM is adopted to generate multi-beam reflective pattern.
The State $0$ unit is shown in Fig. \ref{fig:7}(b),
and the State $1$ unit is formed by rotating the State 0 unit around the $z$-axis with $\ang{90}$.
The conductor patches are $0.6$ mm above the metallic ground.
We use the Particle Swarm Optimization (PSO) algorithm \cite{1997_PSO} to design the codebook to implement a multi-beam reflective pattern.
The EM wave normally incidents to the $32\times{}32$ array, and the codebook is designed to obtain multibeam towards $(\ang{30}, \ang{0})$ and $(\ang{0}, \ang{30})$ at $85$ GHz.

\begin{figure}[htbp]
	\centering
	\vspace{-0.1in}
	\subfloat[]{\includegraphics[width=1.7in]{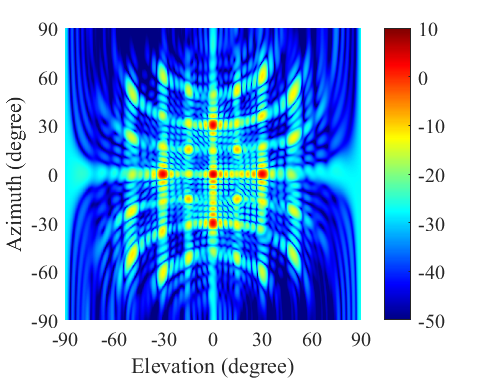}\label{fig:10:a}}
	\hspace{0.0in}
	\subfloat[]{\includegraphics[width=1.7in]{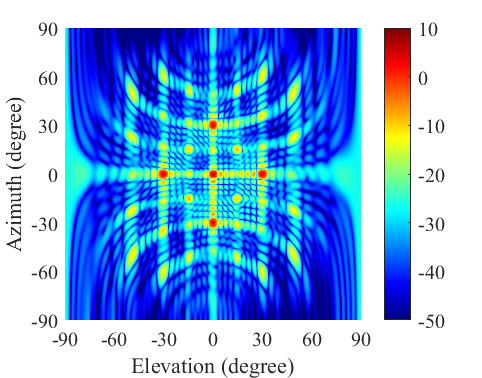}\label{fig:10:b}}
	\caption{2-D reflective pattern of the $32\times{}32$ type-B DCM. (a) Results from R3M. (b) Results from MLFMA. (unit: dBsm)}
	\label{fig:10}
	\vspace{-0.1in}
\end{figure}

Each macro unit of the type-B DCM is discretized with $954$ bases.
The 2-D pattern from R3M-HPE agrees well with that from MLFMA as shown in Fig. \ref{fig:10}.
The beam showing up at $(\ang{0},\ang{0})$ is due to the 
specular reflection,
and the beams at $(\ang{-30},\ang{0})$ and $(\ang{0},\ang{-30})$ are due to the strong gating lobe effects caused by 1-bit quantization.

\subsection{Convergence}
A $16\times{}16$ type-A DCM is utilized to check the convergence.
In this case, 
the TM wave incidents from $(\ang{45}, \ang{0})$, 
and the codebook is designed to reflect the wave towards $(\ang{-30}, \ang{0})$.
The relative residuals evaluated from R3M-HPE at each iteration step with and without preconditioning are compared with those from the classical  $\mathcal{H}$-matrix.

\begin{figure}[htbp]
	\centerline{\includegraphics[width=3.2in]{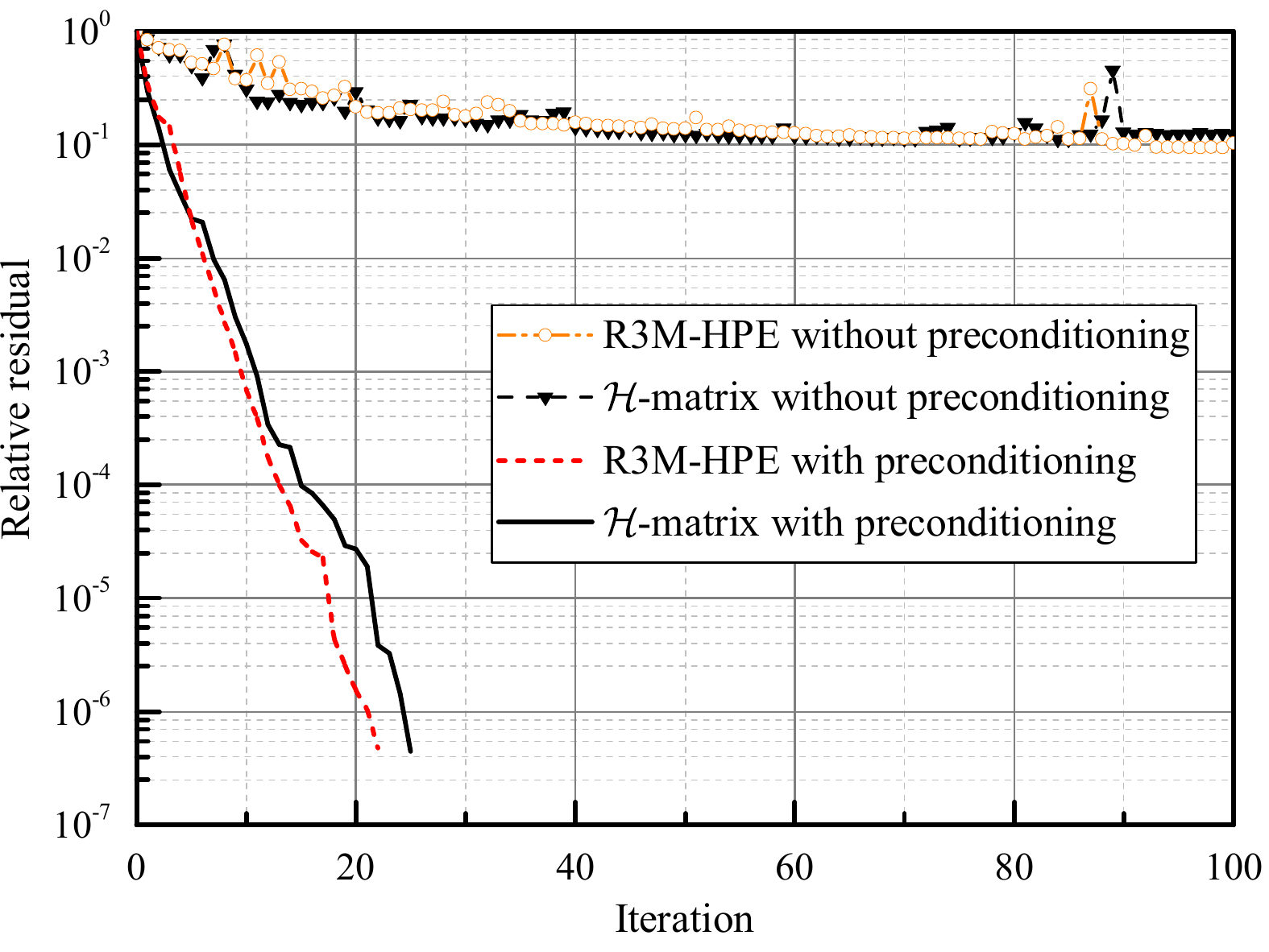}}
	\vspace{-0.0in}
	\caption{The relative residual of iterative solutions with respect to each iteration.}
	\vspace{-0.1in}
	\label{fig:11}
\end{figure}

As shown in Fig. \ref{fig:11}, 
the iterative process of BiCGStab are far from convergence for both R3M-HPE and $\mathcal{H}$-matrix within $100$ iterations without preconditioning.
Unsurprisingly, both converges very fast with the near-field $\mathcal{H}$-LU preconditioning.
R3M-HPE reaches the tolerance of $10^{-6}$ at the $22$-th iteration, 
while the $\mathcal{H}$-matrix does at the $25$-th iteration.
Though the implicit retrieval of the global matrix differs from the direct assembly of $\mathcal{H}$-matrix, 
the process of convergence of R3M-HPE resembles that of $\mathcal{H}$-matrix.


\begin{figure}[ht]
	\vspace{-0.in}
	\centerline{\includegraphics[width=3.2in]{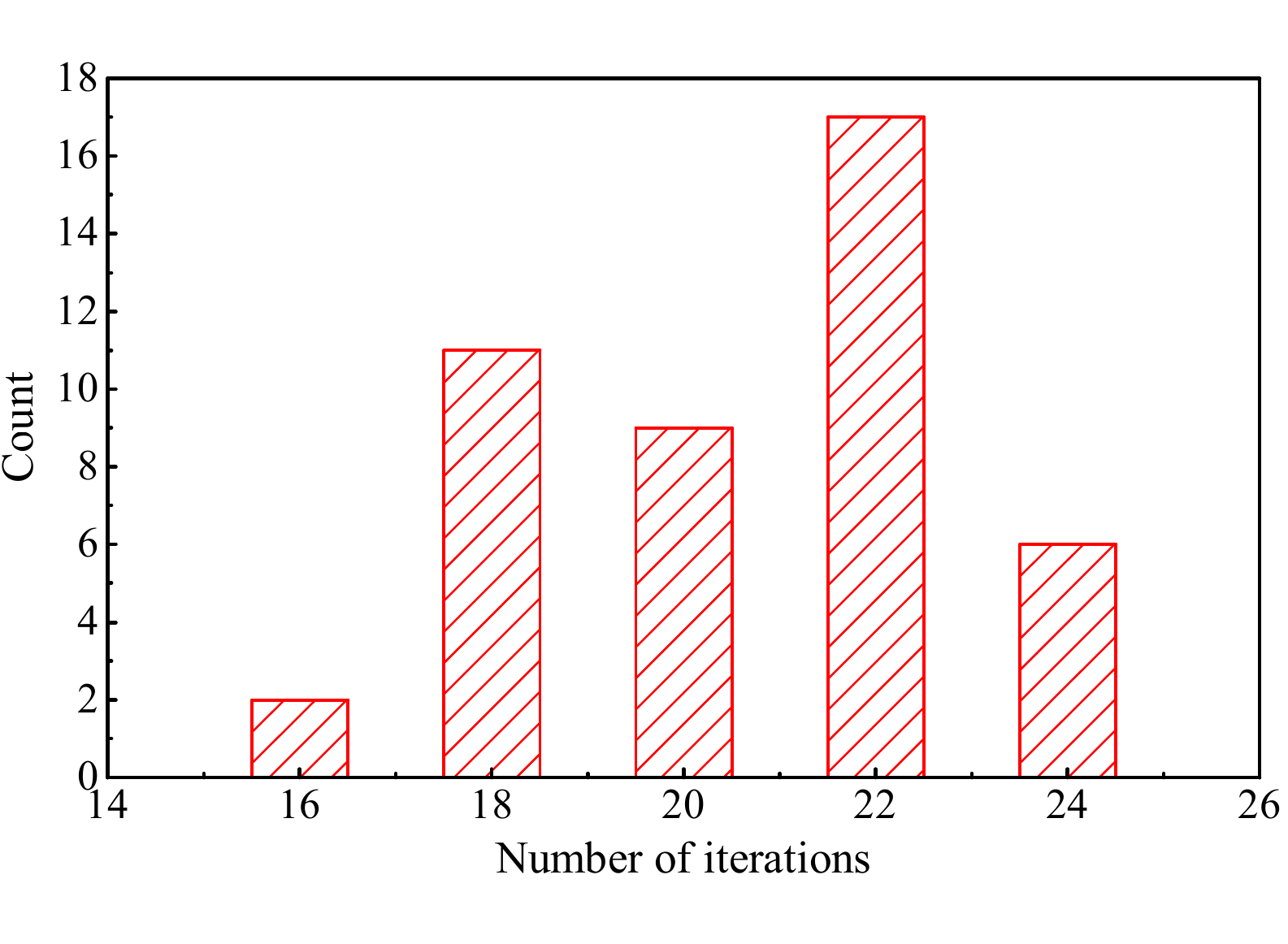}}
	\vspace{-0.1in}
	\caption{Histogram of the number of iterations meeting the preset tolerance for $90$ independent sampling of the codebooks.}
	\vspace{-0.1in}
	\label{fig:12}
\end{figure}

Moreover, we study the convergence of R3M-HPE for the the $16\times{}16$ DCMs with random codebooks.
The principle of randomness is that the states of all the coding units are independent and identically distributed and the state of each unit follows the Bernoulli distribution, which means the probability of each unit being State $0$ is $p$ chosen from $\{0.1,\cdots,0.9\}$.
The sampling of DCMs for each $p$ is repeated $10$ times, which eventually results in $90$ samples.
The histogram of the number of iterations that meet the tolerance of $10^{-6}$ in BiCGStab for each sample is shown in Fig. \ref{fig:12}.
The results demonstrate that R3M-HPE shows good convergence with the near-field $\mathcal{H}$-LU preconditioner.
Since the implicit retrieval of the global matrix via R3M is equivalent to the reconstruction of system matrix from scratch as in (\ref{eq:3}), 
the convergence is justifiably similar to that of the original problem.

\subsection{Computational Efficiency}

\begin{figure}[htbp]
	\subfloat[]{\centerline{\includegraphics[width=3.2in]{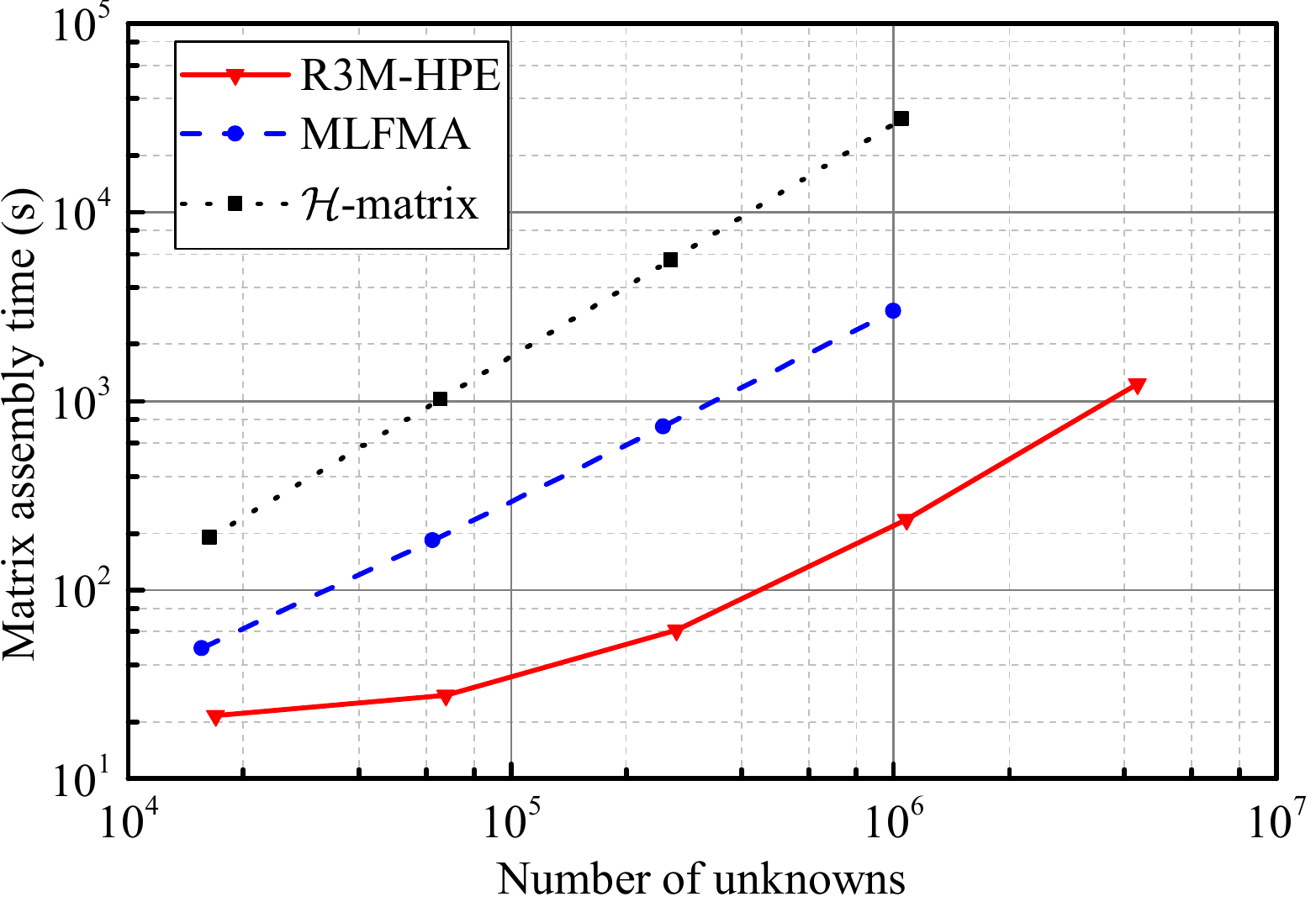}}\label{fig:13:a}}
	\vspace{0.0in}
	\subfloat[]{\centerline{\includegraphics[width=3.2in]{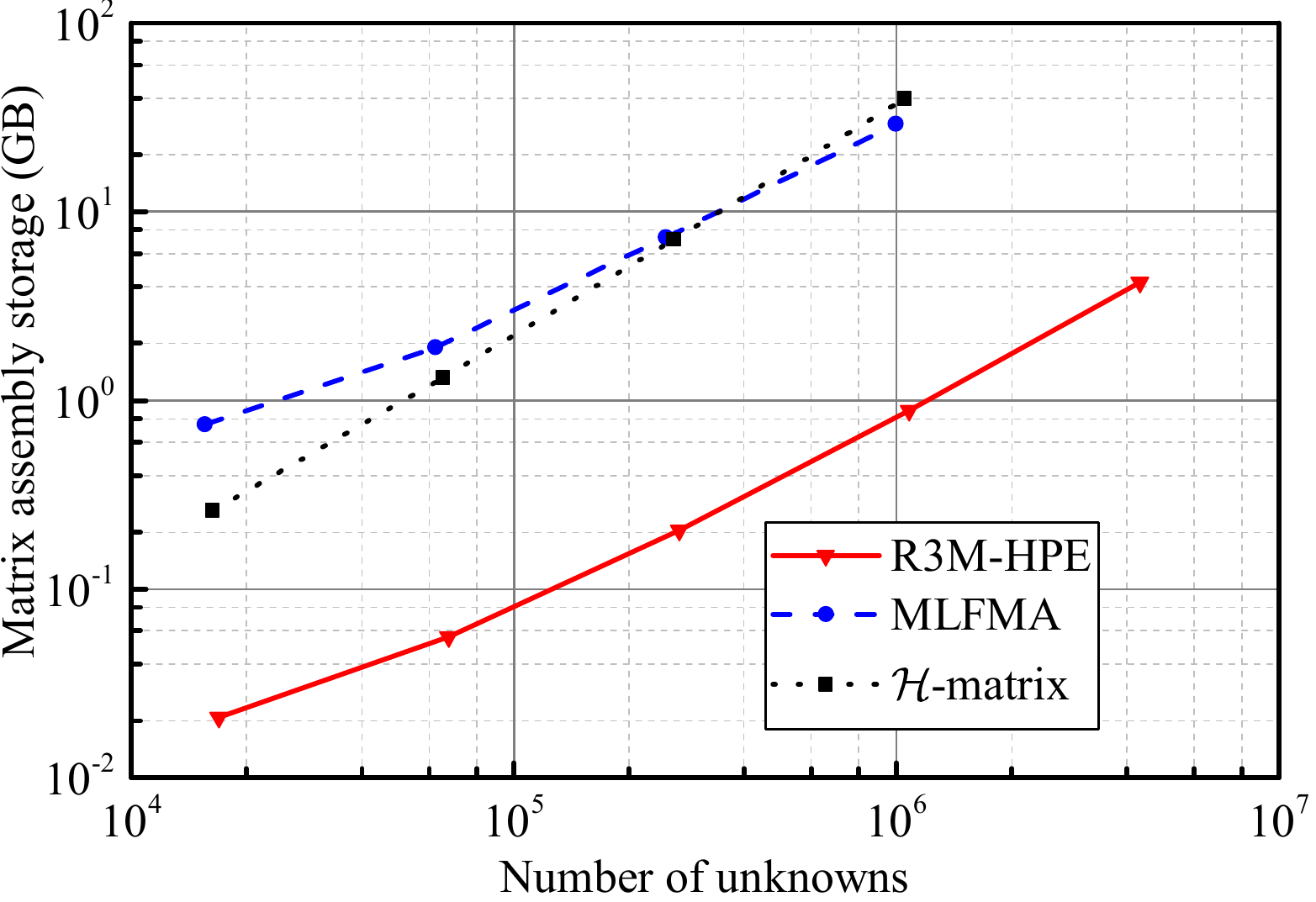}}\label{fig:13:b}}
	\vspace{-0.0in}
	\caption{Computational performance for DCMs with different sizes. (a) Storage for matrix assembly. (b) CPU time for matrix assembly.}
	\vspace{-0.0in}
	\label{fig:13}
\end{figure}

\begin{table*}[!t]
	\centering
	\vspace{-0.0cm}
	\caption{Performance comparison between MLFMA and R3M for DCMs with varied sizes.}
	\vspace{-0.0cm}
	\begin{tabular}{ | c | c | c | c | c | c | c | c | c | c | c | c | c |}
		\hline
		\multirow{3}{*}{\makecell[c]{Array\\size}} & \multicolumn{6}{c|}{MLFMA} & \multicolumn{6}{c|}{R3M-HPE}\\
		\cline{2-13}
		& \multirow{2}{*}{\makecell[c]{Unknowns}} & 
		\multirow{2}{*}{\makecell[c]{Iter.}} & 
		\multirow{2}{*}{\makecell[c]{Solution\\time}} & 
		\multirow{2}{*}{\makecell[c]{Total\\time}} &
		\multirow{2}{*}{\makecell[c]{Solution\\storage}} &
		\multirow{2}{*}{\makecell[c]{Total\\storage}} &
		\multirow{2}{*}{\makecell[c]{Unknowns}} & 
		\multirow{2}{*}{Iter.} & 
		\multirow{2}{*}{\makecell[c]{Solution\\time}} & 
		\multirow{2}{*}{\makecell[c]{Total\\time}} &
		\multirow{2}{*}{\makecell[c]{Solution\\storage}} &
		\multirow{2}{*}{\makecell[c]{Total\\storage}}\\
		& & & & & & & & & & & &\\
		\hline
		$4\times{}4$ &$15,557$& $12$ & $10$ s & $59$ s & $0$ GB & $0.74$ GB &$16,896$&$14$ & $13$ s & $34$ s & $0.07$ GB & $0.09$ GB\\
		\hline
		$8\times{}8$ &$62,383$& $20$ & $28$ s & $211$ s & $1.07$ GB & $2.97$ GB &$67,584$&$21$ & $63$ s & $80$ s & $0.29$ GB & $0.34$ GB\\
		\hline
		$16\times{}16$ &$249,685$& $31$ & $157$ s & $981$ s & $4.7$ GB & $12.0$ GB &$270,336$& $30$ & $298$ s & $359$ s & $1.17$ GB & $1.38$ GB\\
		\hline
		$32\times{}32$ &$998,588$& $47$ & $1276$ s & $4645$ s & $17.5$ GB & $46.7$ GB &$1,081,344$& $50$ & $1413$ s & $1649$ s & $4.76$ GB & $5.65$ GB\\
		\hline
		$64\times{}64$ & \multicolumn{6}{c|}{Out of memory} &$4,325,376$& $83$ & $6932$ s & $7167$ s & $19.07$ GB & $23.28$ GB\\
		\hline		
	\end{tabular}
	\vspace{-0.0cm}
	\label{tab:1}
\end{table*}

R3M-HPE has advantages in simulating a series of DCMs with varied codebooks.
The method should firstly be efficient for each one-shot simulation.
We still take the type-A DCM as an example to verify the efficiency of R3M-HPE.
The discretization of the macro unit is the same as the previous examples.
The array sizes of the DCMs in this example are $4\times{}4$, $8\times{}8$, $16\times{}16$, $32\times{}32$, and $64\times{}64$.
The excitation is fixed as the TM-polarized incident wave from $(\ang{45}, \ang{0})$.
The codebooks are designed to reflect the waves towards $(\ang{-30}, \ang{0})$ for all the DCMs with different sizes.

The storage and CPU time cost of R3M-HPE needed in analyzing the DCMs are compared with those of the classical $\mathcal{H}$-matrix and MLFMA.
The number of unknowns in R3M-HPE is slightly larger than those of the classical $\mathcal{H}$-matrix and MLFMA,
due to the extended bases at the boundaries.
In terms of the low-rank compression, 
R3M-HPE keeps the precision the same as $\mathcal{H}$-matrix,
since the virtual $\mathcal{H}$-matrix resulted from HPE is the memory-shared version of $\mathcal{H}$-matrix.
As shown in Fig. \ref{fig:13}, the storage and CPU time cost for matrix assembly of R3M-HPE is $1$ to $2$ orders of magnitude less than those of the classical $\mathcal{H}$-matrix, exhibiting significant improvements within the same paradigm of calculation.
Furthermore, R3M-HPE also substantially exceeds MLFMA in both the storage and CPU time cost.
The CPU time of R3M-HPE to assemble the matrix for the $32\times{}32$ and $64\times{}64$ is approximately $1/10$ that of MLFMA, as shown in Fig. \ref{fig:13}(a).
In terms of storage, MLFMA takes $10$ times more memory for matrix assembly than R3M-HPE for cases in this example.
Besides, the required storage of MLFMA for the $64\times{}64$ array exceeds the RAM limit, while R3M-HPE can still be executed with around $5$ GB memory cost, enabling the analysis of larger size of DCMs.


\begin{figure}[htbp]
	\centering
	\includegraphics[width=3.4in]{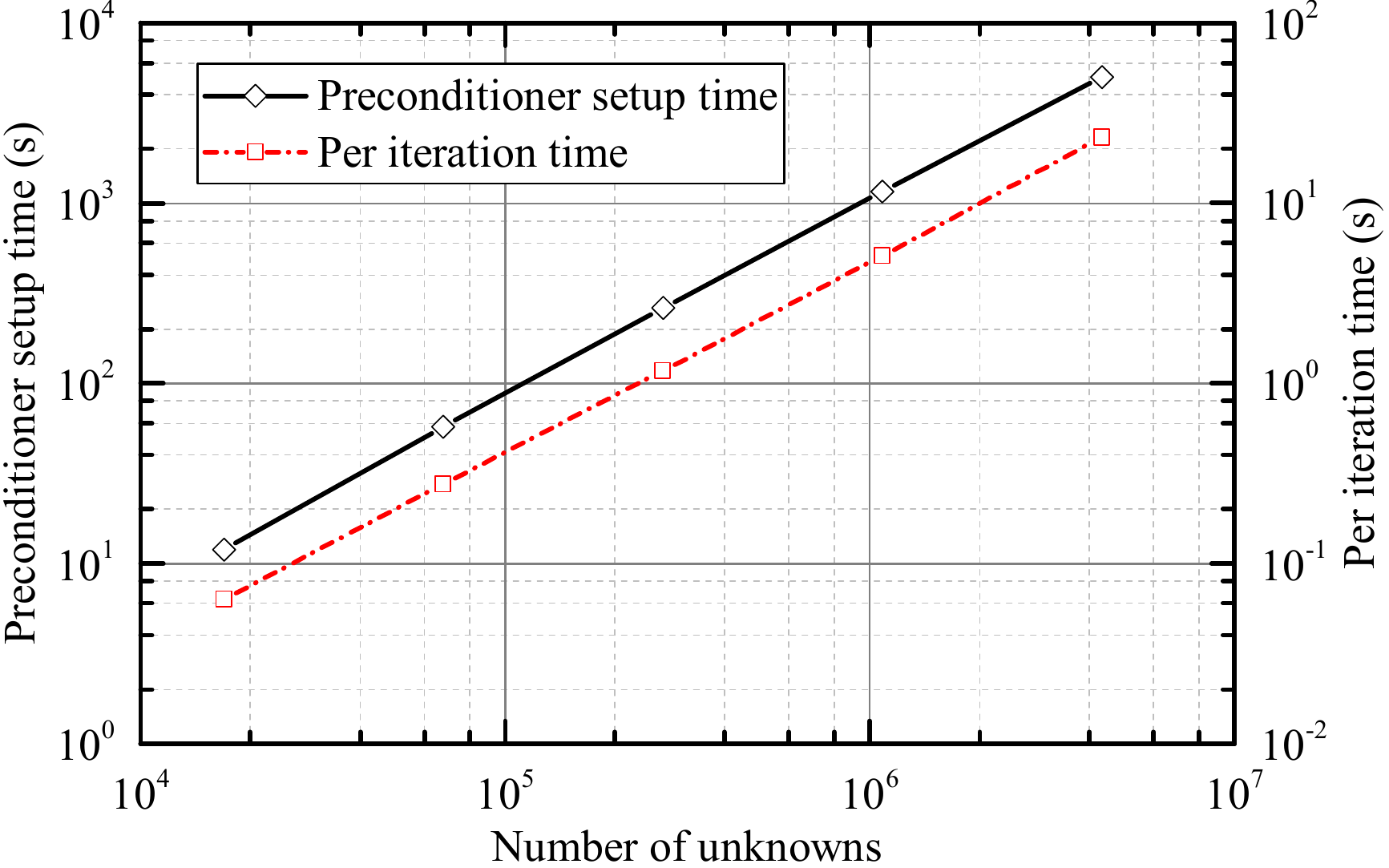}
	\vspace{0.0in}
	\caption{CPU time cost of the iterative solution for DCMs with varied sizes.}
	\vspace{0.0in}
	\label{fig:14}
\end{figure}

The setup time of the preconditioner and per iteration time during the solution of R3M-HPE are shown in Fig. \ref{fig:14}.
The detailed information on the cost of the iterative solutions are shown in Table \ref{tab:1}.
As shown, the number of iterations (abbreviated as Iter.) are close to each other, 
whereas the solution CPU time of R3M-HPE is relatively higher than that of MLFMA.
Nonetheless, the total CPU time cost of R3M-HPE involving both the matrix assembly and the iterative solution is still much less than that of MLFMA.
The storage cost during the solution, 
for both methods increases approximately linearly with the number of unknowns.
As is observed, the solution storage of R3M-HPE is also $3$ to $4$ times less.
Consequently, the total storage cost is much less than that of MLFMA.
Note that R3M-HPE takes a relatively small proportion of storage compared with the preconditioner, 
because the global matrix is stored in the form of virtual $\mathcal{H}$-matrix.
Though the advantage of R3M-HPE in storage is slightly weakened by the preconditioner in the iterative solution, 
the all-inclusive cost still outperforms MLFMA.
Additionally, R3M-HPE would still save approximately $50\%$ more storage than MLFMA if the same preconditioner is adopted.
The superior performance of R3M-HPE in less storage and CPU time cost stems from the pruning of redundant computations in HPE, 
which fully exploits the rigid periodicity resulted from R3M. 

To demonstrate the performance of R3M-HPE for the fast validation of a series of DCMs, 
we design dedicated codebooks for $25$ different targets, 
which is more practical than random codebooks.
The TM wave still incidents from $(\ang{45}, \ang{0})$.
The sets of values for azimuth and elevation are both $\{\ang{-30}, \ang{-15}, \ang{0}, \ang{15}, \ang{30}\}$, leading to $25$ combinations.
The size of the DCM is fixed as $32\times{}32$.


\begin{table}[htbp]
	\centering
	\vspace{-0.0cm}
	\caption{Performance comparison between MLFMA and R3M-HPE for fast validations of a group of DCMs.}
	\vspace{-0.0cm}
	\begin{tabular}{ | c | c | c | c |}
		\hline
		\multicolumn{2}{c|}{MLFMA} & \multicolumn{2}{|c|}{R3M-HPE}\\
		\hline
		Peak storage & Total CPU time & Peak storage & Total CPU Time\\
		\hline
		$53.9$ GB & $31$ h $14$ min & $5.59$ GB & $9$ h $42$ min  \\
		\hline
	\end{tabular}
	\vspace{-0.0cm}
	\label{tab:2}
\end{table}

\begin{figure}[htbp]
	\centering
	\subfloat[]{\includegraphics[width=1.7in]{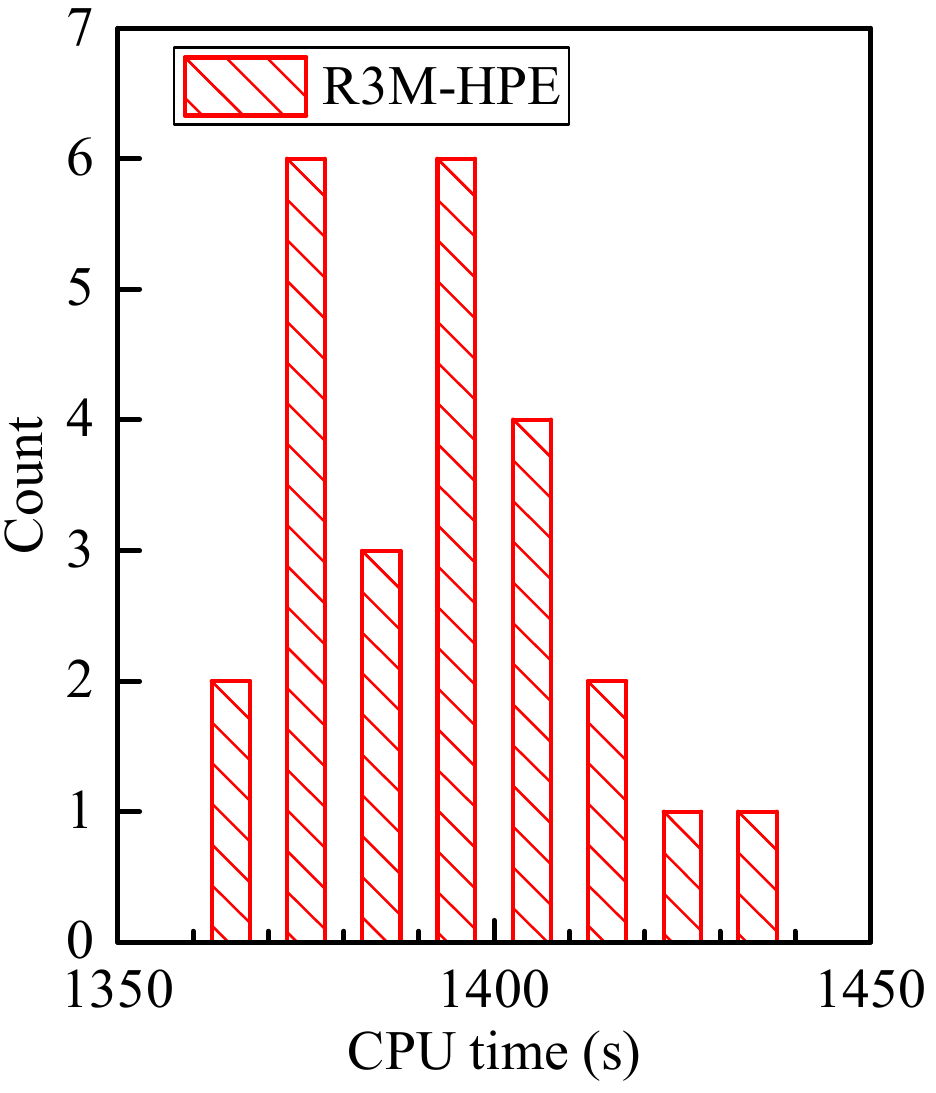}\label{fig:14:a}}
	\hspace{0.0in}
	\subfloat[]{\includegraphics[width=1.7in]{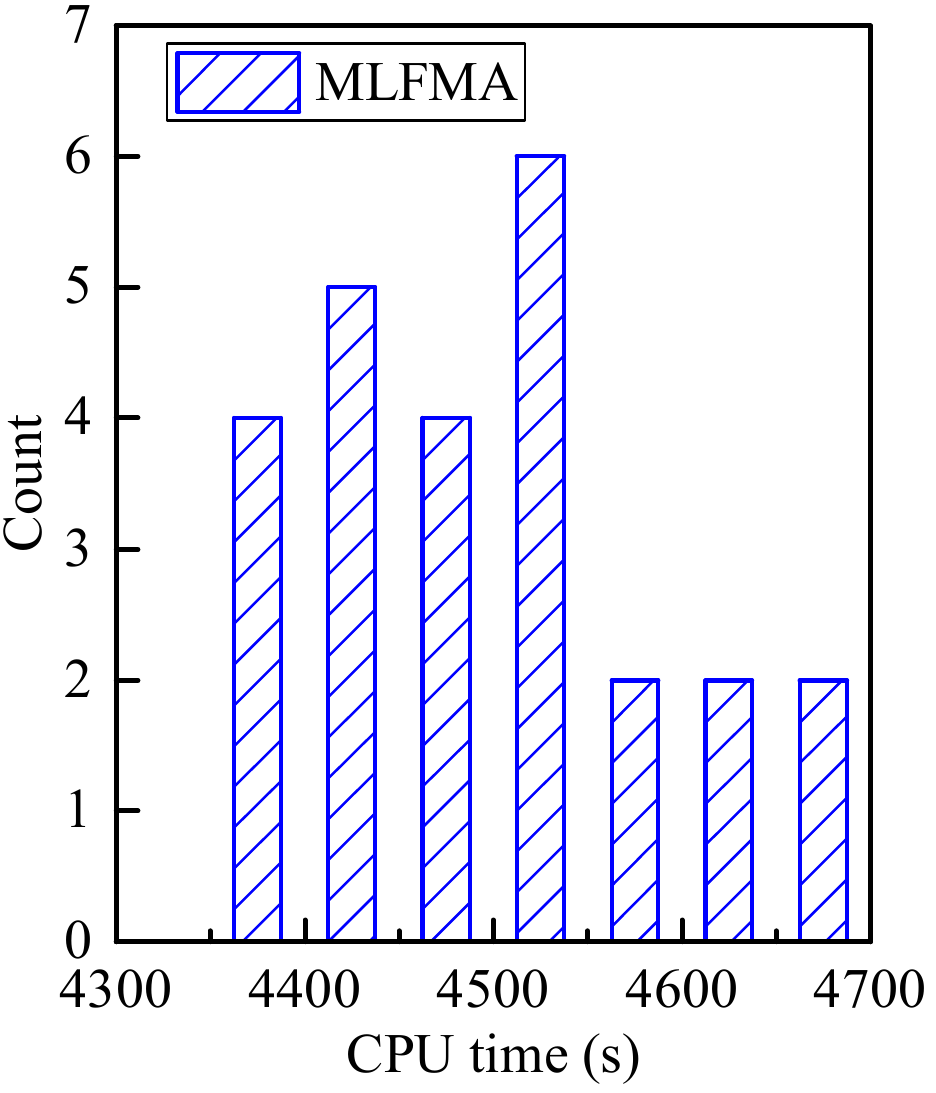}\label{fig:14:b}}
	\vspace{0.0in}
	\caption{Histogram of the CPU time cost for the $25$ codebooks. Storage. (a) Results from R3M-HPE. (b) Results from MLFMA.}
	\vspace{0.0in}
	\label{fig:15}
\end{figure}

The peak storage and total CPU time cost of fast validations of the $25$ DCMs are shown in Table \ref{tab:2}.
It is rational that the peak storage for all cases is close to that in Table \ref{tab:1},
and the CPU time are close to $25$ times the total CPU time in Table \ref{tab:1}.
R3M-HPE takes less than $1/9$ the storage and $1/3$ CPU time of MLFMA for the multi-round validations.
Since the global matrix is built only once, the average CPU time cost is further reduced in R3M-HPE.
The histogram of the CPU time cost for the $25$ codebooks is shown in Fig. \ref{fig:15}.
The CPU time cost of R3M-HPE for validating each DCM ranges from $1360$ s to $1440$ s,
while for MLFMA it ranges from $4350$ s to $4700$ s,
indicating that R3M-HPE consistently outperforms MLFMA in all cases.
In other words, R3M-HPE is effective and efficient for both the single and multi-round fast validations of DCMs.

\section{Conclusion}
\label{sec:conclustion}
We propose the recurrence rebuild and retrieval method for the fast electromagnetic validations of large-scale DCMs.
This method enables the construction of a global matrix that corresponds to a rigorously periodic array containing all possible codebooks.
The global matrix is assembled only once and shared to match arbitrary codebooks via implicit retrieval.
The global matrix from R3M could be even stored offline in preparation for matching any potential codebook. 
By leveraging the hierarchical pattern exploitation algorithm, the assembly of the global matrix is further accelerated.
Numerical examples show that 
R3M-HPE achieves 1 to 2 orders of magnitude less storage and CPU time for matrix assembly than the classical $\mathcal{H}$-matrix. 
Moreover, R3M-HPE also outperforms MLFMA in efficiency for DCMs with the array size ranging from $4\times{}4$ to $64\times{}64$.
For multi-round validations, R3M-HPE takes less than $1/9$ storage and $1/3$ CPU time of MLFMA for $25$ practical codebooks, 
which indicates the high efficiency of the proposed method.

\bibliography{IEEEabrv}
\bibliographystyle{IEEEtran}


\end{document}